\PassOptionsToPackage{unicode}{hyperref}
\PassOptionsToPackage{hyphens}{url}
\documentclass[
  12pt,
]{article}
\usepackage{scicite}

\usepackage{times}

\usepackage{mathtools}
\usepackage{amsmath}
\usepackage{amsthm}
\usepackage{amssymb}
\usepackage[italicdiff]{physics}
\mathtoolsset{showonlyrefs}

\usepackage[lmargin=0.75in,rmargin=0.75in,tmargin=0.75in,bmargin=0.75in]{geometry}

\let\citep\cite
\let\citet\cite
\IfFileExists{bookmark.sty}{\usepackage{bookmark}}{\usepackage{hyperref}}
\IfFileExists{xurl.sty}{\usepackage{xurl}}{}
\hypersetup{
  pdftitle={Evaluating Bias and Noise from Census Privacy Methods},
  pdfauthor={Christopher T. Kenny; Cory McCartan; Shiro Kuriwaki; Tyler Simko; Kosuke Imai},
  hidelinks,
  pdfcreator={LaTeX via pandoc}}

\usepackage{longtable,booktabs,array}
\usepackage{calc} 
\usepackage{etoolbox}
\makeatletter
\patchcmd\longtable{\par}{\if@noskipsec\mbox{}\fi\par}{}{}
\makeatother
\IfFileExists{footnotehyper.sty}{\usepackage{footnotehyper}}{\usepackage{footnote}}
\makesavenoteenv{longtable}
\usepackage{graphicx}
\makeatletter
\def\maxwidth{\ifdim\Gin@nat@width>\linewidth\linewidth\else\Gin@nat@width\fi}
\def\maxheight{\ifdim\Gin@nat@height>\textheight\textheight\else\Gin@nat@height\fi}
\makeatother
\setkeys{Gin}{width=\maxwidth,height=\maxheight,keepaspectratio}
\makeatletter
\def\fps@figure{htbp}
\makeatother

\theoremstyle{plain}
\newtheorem{prop}{Proposition}[section]

\newcommand{\Z}{\ensuremath{\mathbb{Z}}}

\newcommand{\eps}{\varepsilon}

\DeclareMathOperator{\E}{\mathbb{E}}

\DeclareMathOperator{\Var}{\mathbb{V}}
\DeclareMathOperator{\Cov}{Cov}

\newcommand{\Norm}{\mathcal{N}\qty}
\newcommand{\MVNorm}[1][]{\mathcal{N}_{#1}\qty}

\newcommand{\pp}{\mathrm{td}}
\newcommand{\td}{\mathrm{td}}
\newcommand{\nm}{\mathrm{nm}}
\newcommand{\sw}{\mathrm{sw}}
\newcommand{\I}{\mathcal{I}}
\newcommand{\St}{\mathcal{S}}
\newcommand{\aian}{\textsc{ai/an}}
\newcommand{\mse}{\textsc{mse}}
\DeclareMathOperator{\diag}{diag}

\usepackage{placeins}

\makeatletter
\@ifpackageloaded{caption}{}{\usepackage{caption}}
\AtBeginDocument{%
\ifdefined\contentsname
  \renewcommand*\contentsname{Table of contents}
\else
  \newcommand\contentsname{Table of contents}
\fi
\ifdefined\listfigurename
  \renewcommand*\listfigurename{List of Figures}
\else
  \newcommand\listfigurename{List of Figures}
\fi
\ifdefined\listtablename
  \renewcommand*\listtablename{List of Tables}
\else
  \newcommand\listtablename{List of Tables}
\fi
\ifdefined\figurename
  \renewcommand*\figurename{Figure}
\else
  \newcommand\figurename{Figure}
\fi
\ifdefined\tablename
  \renewcommand*\tablename{Table}
\else
  \newcommand\tablename{Table}
\fi
}
\@ifpackageloaded{float}{}{\usepackage{float}}
\floatstyle{ruled}
\@ifundefined{c@chapter}{\newfloat{codelisting}{h}{lop}}{\newfloat{codelisting}{h}{lop}[chapter]}
\floatname{codelisting}{Listing}

\makeatother
\makeatletter
\@ifpackageloaded{caption}{}{\usepackage{caption}}
\@ifpackageloaded{subcaption}{}{\usepackage{subcaption}}
\makeatother

\newenvironment{sciabstract}{%
\begin{quote} \bf}
{\end{quote}}

\title{\vspace*{-1.5em}Evaluating Bias and Noise Induced by the U.S.
Census Bureau's Privacy Protection Methods}

\author{%
Christopher T. Kenny,$^{1\dagger}$
Cory McCartan,$^{2\dagger}$
Shiro Kuriwaki,$^{3}$
Tyler Simko,$^{1}$
Kosuke Imai$^{1,4\ast}$
\\
\\
\normalsize $^{1}$%
Department of Government, Harvard University,\\
\normalsize 1737 Cambridge Street, Cambridge, MA 02138
\\
\normalsize $^{2}$%
Center for Data Science, New York University
\\
\normalsize $^{3}$%
Department of Political Science, Yale University
\\
\normalsize $^{4}$%
Department of Statistics, Harvard University
\\%
\\\normalsize $^\ast$To whom correspondence should be addressed; E-mail:  imai@harvard.edu.
\\\normalsize $^\dagger$These authors contributed equally to this work.
}

\date{}

\begin{document}

\baselineskip16pt

\maketitle

\begin{sciabstract}
The United States Census Bureau faces a difficult trade-off between the
accuracy of Census statistics and the protection of individual
information. We conduct the first independent evaluation of bias and
noise induced by the Bureau's two main disclosure avoidance systems: the
TopDown algorithm employed for the 2020 Census and the swapping
algorithm implemented for the three previous Censuses. Our evaluation
leverages the Noisy Measurement File (NMF) as well as two independent
runs of the TopDown algorithm applied to the 2010 decennial Census. We
find that the NMF contains too much noise to be directly useful without
measurement error modeling, especially for Hispanic and multiracial
populations. TopDown's post-processing reduces the NMF noise and
produces data whose accuracy is similar to that of swapping. While the
estimated errors for both TopDown and swapping algorithms are generally
no greater than other sources of Census error, they can be relatively
substantial for geographies with small total populations.
\end{sciabstract}

\vspace*{1em}
\noindent\textbf{\textit{Teaser}}\quad Old and new Census Bureau privacy
protection systems introduce generally small errors, except for less
populated geographies.

\newpage

\section{Introduction}\label{sec-intro}

Population statistics produced by the U.S. Census serve as the basis for
many consequential public policy decisions, such as legislative
redistricting and the disbursement of over \$675 billion in federal
funds \citep{funds2017, kenny2021impact}. At the same time, the U.S.
Census Bureau is legally obligated to protect the confidentiality of
individual responses to the Census questionnaires (US Code Title 13).
This means that the Bureau faces a difficult trade-off between data
accuracy and privacy protection.

The Bureau introduced a new disclosure avoidance system (DAS) for the
2020 decennial Census named TopDown, based on a mathematical definition
of privacy called \emph{differential privacy} \citep{topdown}. The
TopDown algorithm first injects mean-zero, independent random noise to
tabulations of the confidential data set (the Census Edited File or
CEF), which contains individual information about every enumerated
resident in the United States. The resulting Noisy Measurement File
(NMF) is then post-processed to ensure that the final census data
satisfy a set of requirements for straightforward data use; for example,
all the resulting tabulations are non-negative integers and are
consistent across different census geographic levels.

The TopDown algorithm in the 2020 decennial Census replaced the old DAS,
which was based on a \emph{swapping} algorithm and used in the 1990,
2000, and 2010 Censuses \citep{mckenna2018disclosure}. The swapping
algorithm randomly selects a household record in a small census block
and interchanges it with a similar record in another block. Households
with higher disclosure risk, because of their unique attributes or the
small population of their block, are swapped with higher probability.

This change to the DAS led to debates among scholars and stakeholders
regarding the appropriate balance between data accuracy and privacy
protection
\citep{ruggles19, santos20, kenny2021impact, ncai2020, scariano2022balancing, hotzpnas, Hotz2022Chronicle, ruggles22}.
While the Bureau has released a series of demonstration datasets and
error metrics of the TopDown algorithm, the NMF was not released until
April 2023 \citep{census2023may, nmflawsuit}. The Bureau also has yet to
release metrics that directly compare the TopDown error to the NMF and
swapping algorithm errors.

We provide the first independent assessment of the bias and noise
induced by the TopDown algorithm and the swapping algorithm, relative to
the CEF. Although we cannot directly observe the CEF, we use the known
statistical properties of the noise added to the CEF to evaluate bias
and uncertainty of the entire TopDown algorithm. Averaging the
discrepancies between the TopDown and NMF statistics over a set of
geographies forms an unbiased estimator of the TopDown bias, because the
NMF is an unbiased estimate of the CEF itself. We compare the bias and
noise caused by the swapping algorithm, the TopDown algorithm, and the
noise induced in the NMF prior to post-processing. Errors from TopDown
can be put in context with the estimated sizes of coverage and
non-sampling errors by Census researchers \citep{memo21}.

Because this approach is fundamentally statistical, we are limited to
estimating the \emph{average bias}, i.e., the errors in the counts
averaged over thousands or millions of census geographies. Such
aggregation may mask biases that potentially exist in some census
geographies. However, our estimates of the TopDown and swapping noise,
as measured by their \emph{root mean square error}, capture important
distributional information about these errors that can be useful to
practitioners in evaluating the reliability of census statistics. We
also study how the noise and bias depend on the overall population of
census geographies, providing additional insight into how the
DAS-induced errors may vary across these geographies.

Our evaluation is possible because of the fortuitous availability of
three distinct data sources based on the 2010 decennial Census. The
first is the NMF and corresponding TopDown output (including
post-processing), contained in the 2010 Redistricting and DHC Production
Settings Demonstration Data published in April 2023 \citep{nmf2010}.
Second, we use the official release of the 2010 decennial census, which
represents the output of the swapping algorithm. We transform the
information in the NMF to match the traditional Census geographies in
the official census. Third, we also use another 2010 Demonstration Data
(June 2021 vintage) released in August 2021 \citep{census2021aug}. This
demonstration data is based on a different run of the TopDown algorithm,
providing us with an additional independent output of the new DAS (The
Bureau had initially refused to release the NMF but recently made it
publicly available. Unfortunately, the Bureau deleted the NMF used to
generate the August 2021 Demonstration Data \citep{foialetter}. This
means that the April 2023 Demonstration Data are based on a new,
independent run of the TopDown algorithm). Since these two releases are
statistically independent, we can use them alongside known statistical
properties of the noise-injection stage of the TopDown algorithm to
quantify the uncertainty of our estimates. The details of our
methodology are described in Materials and Methods
(Section~\ref{sec-addl}).

Our key contribution is to evaluate the bias and noise of the TopDown
algorithm relative to the CEF: the non-privatized microdata. Prior work
attempting to evaluate the bias induced by the TopDown algorithm has
been limited to applying an experimental version of differential privacy
\citep{Asquith2022Assessing}, often to simulated data the researcher
creates \citep{christ2022differential}. Other work compares the TopDown
algorithm against the swapping algorithm only
\citep{cohen_et_al, kenny2021impact}. The methodology we develop may
therefore also be useful for practitioners who wish to characterize the
bias and noise of particular published statistics beyond those
considered here.

Our findings, detailed in the Results section and are based on the
methods described in the Materials and Methods section, can be
summarized as follows. Overall, both the TopDown and swapping algorithms
produce nearly-unbiased population and racial group counts, on average.
Variance is generally small in absolute terms for most geographies, but
errors are substantially larger for certain racial/ethnic groups,
particularly for Hispanic and multiracial populations. TopDown's
post-processing reduces the noise in the NMF, and ultimately the
variance from swapping appears to be roughly similar to the variance
from TopDown. These patterns hold across census geographies with varying
population sizes and racial diversity. Generally, the TopDown and
swapping errors are no greater than other sources of errors, such as
undercounting and missing responses. For small-population geographies,
errors from both TopDown and swapping can be large relative to the total
population as well as other sources of error, warranting caution in
conducting statistical analyses which rely on these small counts.

\section{Materials and Methods}\label{sec-addl}

In this section, we describe our methodology used to evaluate the bias
and uncertainty of the TopDown and swapping algorithms. This methodology
may be useful for other researchers who wish to quantify the average
bias and variance of a set of statistics published after TopDown or
swapping have been applied.

\subsection{Census and NMF
Geographies}\label{census-and-nmf-geographies}

\begin{figure}[t]

\centering{

\includegraphics[width=\textwidth,height=0.5\textheight]{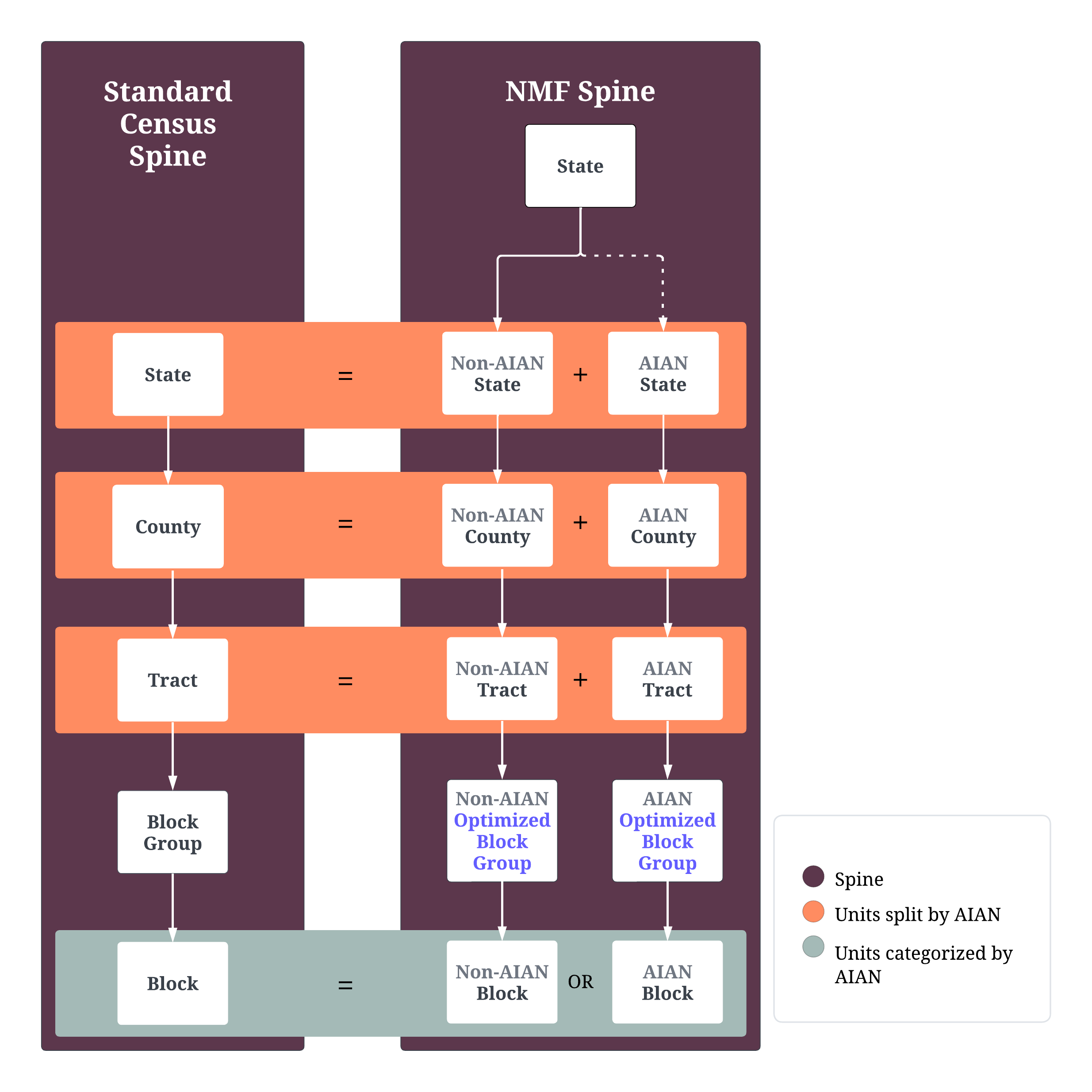}

}

\caption{\label{fig-nmf-heirarchy}\textbf{The Census geographic
hierarchy (spines).} Spines for standard census geographies and for the
hierarchy in the Noisy Measurement File (NMF). Higher units indicate
enclosing units. In the right spine, sub-state geographic units are
split into American Indian / Alaska Native (\aian{}) and non-\aian{}
portions. The dotted arrow indicates that not all states have an \aian{}
segment. The teal area indicates that the units can be matched
one-to-one across spines. For example, a single block is never split
into a Non-\aian{} vs.~\aian{} fragment, so a block from the NMF spine
matches 1:1 to a block in the standard spine.}

\end{figure}

The Census Bureau partitions the entire geographic area of the United
States into \emph{census blocks}, of which there were 11,078,297 in
2010. As shown in Figure~\ref{fig-nmf-heirarchy}, sets of blocks can be
hierarchically grouped together to form a \emph{geographic spine}. The
most familiar geographic spine is the \emph{standard Census spine},
where blocks belong to block groups, which in turn belong to tracts,
counties, and states. The NMF is built around a slightly different
\emph{NMF spine}.

Figure~\ref{fig-nmf-heirarchy} displays the census spines for the
standard census geographies and for the NMF geographies. Geographies
larger than blocks are split in the NMF according to whether they
include American Indian / Alaska Native (\aian) areas, in an effort to
improve accuracy in these areas. For example, a tract in a typical
census release is on the left spine, but is broken into \aian{} and
non-\aian{} areas on the right spine. Blocks retain a one-to-one
correspondence, but each is designated as part of the \aian{} or
non-\aian{} spine. Block groups are on-spine within each respective
hierarchy, but they represent different geographies. In other words,
optimized block groups on the NMF spine do not necessarily correspond to
block groups on the traditional census spine. All analyses herein use
the released Block Groups, not Optimized Block Groups. Note that Block
Groups, the geography in Census releases, are distinct from Optimized
Block Groups used in TopDown (see
\url{https://www.census.gov/about/training-workshops/2021/2021-07-01-das-presentation.pdf}
for further details). As a result, we refer to Block Groups as
``off-spine'' for TopDown, even though Block Groups are on the
traditional Census spine.

We transformed the NMF into traditional Census geographies, such as
block groups and voting districts, by building a mapping from the unique
identifiers of NMF on-spine geographies, \emph{geocodes}, to traditional
Census \emph{GEOID} identifiers. We describe this cross-walking process
in detail in Supplementary Materials Section S1.

\subsection{Setup and Notation}\label{sec-setup}

We use \(\Gamma\) to denote the set of all geographies on the NMF
spine---blocks, \aian{} tracts, non-\aian{} counties, etc. The Census
Edited File (CEF) can be equivalently represented as a set of
\emph{histograms}, \(\{\vb x_\gamma\}_{\gamma\in\Gamma}\), one for every
geography \(\gamma\in\Gamma\), which count the number of people in that
geography with every possible combination of measured attributes. For
the redistricting NMF, the total number of bins for each histogram is
\(2016=(2\text{ voting-age categories})\times (2\text{ Hispanic categories})\times (63\text{ race categories})\times (8\text{ housing types})\).

In our analysis, we regard the CEF as the ground truth and estimate
average bias and root mean square error relative to it. The detailed
histograms \(\{\vb x_\gamma\}_{\gamma\in\Gamma}\) can be aggregated to a
set \(\St\) of statistics of interest
\(\{\vb y^*_\gamma\}_{\gamma\in\Gamma}\), like the total voting-age
population, or Black population, using a \(|\St|\times 2016\)
aggregation matrix \(A\): \[\vb y^*_\gamma \coloneqq A\vb x_\gamma.\]
However, simply aggregating the confidential CEF
\(\{\vb x_\gamma\}_{\gamma\in\Gamma}\) does not necessarily provide any
privacy guarantee to respondents. To protect respondent confidentiality,
the Bureau has developed \emph{disclosure avoidance} methods, which
range from simple (suppressing parts of
\(\{\vb y^*_\gamma\}_{\gamma\in\Gamma}\)) to complex. We examine the
TopDown algorithm, developed for the 2020 Census, and compare its
properties with the \emph{swapping} method, used for the 1990, 2000, and
2010 Censuses.

\subparagraph{TopDown Algorithm.}\label{topdown-algorithm.}

To produce a differentially private version of
\(\{\vb y^*_\gamma\}_{\gamma\in\Gamma}\), the Bureau applies the TopDown
algorithm, which proceeds in two main steps.

First, in each geography \(\gamma\in\Gamma\), \emph{noisy measurements}
are made of \(\vb x_\gamma\) through a series of \emph{queries} encoded
in a known query matrix \(Q\):
\[\vb{\widetilde M}_\gamma \coloneqq Q\vb x_\gamma + \vb*\eta_\gamma, \qquad
    \vb*\eta_\gamma \sim \MVNorm[\Z](0, \Sigma_\gamma),\] where
\(\mathcal{N}_\Z\) is the discrete Gaussian distribution, and
\(\Sigma_\gamma=\diag(\vb*\varsigma^2_\gamma)\) is a diagonal covariance
matrix, with each query's variance \(\varsigma^2_{\gamma j}\) determined
by a privacy-loss budget schedule set in advance by the Bureau. It is
this noisy measurement step which provides zero-Concentrated
Differential Privacy (z-CDP) guarantees \citep{canonne2020discrete}.

Second, the noisy measurements are hierarchically post-processed using a
multi-pass optimization and rounding routine. The post-processing
ensures that published statistics are consistent across and within
geographies, and enforces policy constraints. For example, these
constraints ensure that all statistics will be non-negative, and that
the published state population totals exactly match the population
totals from the CEF. \citet{topdown} presents more information on the
post-processing including the justification for these constraints.

For our purposes, we can write the output of the post-processing step as
a new set of noise-infused microdata,
\[\{\vb X^\pp_\gamma \}_{\gamma\in\Gamma} 
    = g(\{(\vb{\widetilde M}_\gamma, \vb x_\gamma) \}_{\gamma\in\Gamma}),\]
where the dependence on the CEF counts \(\vb x_\gamma\) is necessary
because certain constraints are functions of the original microdata. The
new microdata can be aggregated, yielding differentially private
statistics \[\vb Y^\pp_\gamma \coloneqq A\,\vb X^\pp_\gamma\] for every
geography \(\gamma\in\Gamma\). We write the error produced by the
TopDown algorithm as
\[\vb*\eps^\pp_\gamma \coloneqq \vb Y^\pp_\gamma - \vb y^*_\gamma
= A(\vb X^\pp_\gamma - \vb x_\gamma).\]

The post-processed statistics \(\vb Y^\pp\) cannot be directly compared
to the noisy measurements \(\vb{\widetilde M}\), since the latter are at
the query level rather than the statistic level. This means that a
single statistic, like the total population, may be computed separately
from multiple different queries, each producing a different answer.
Fortunately, using the workflow described in \citet{mccartan2023nmf}, it
is possible to combine the noisy measurements from all the queries to
form unbiased, minimum-variance estimates of \(\vb y^*_\gamma\). This is
accomplished by aggregating the noisy measurements to form multiple
noisy estimates of each statistic, and then combining these multiple
estimates with an inverse-variance-weighted mean.

Formally, we define an aggregation-weighting matrix \(B(\Sigma)\) which
performs both these steps at once, yielding
\[\vb Y^\nm_\gamma \coloneqq B(\Sigma_\gamma)\vb{\widetilde M}_\gamma 
    = B(\Sigma_\gamma)Q\vb x_\gamma + B(\Sigma_\gamma)\vb*\eta_\gamma\]
for every geography \(\gamma\in\Gamma\). The noisy measurement error,
\begin{equation}\phantomsection\label{eq-def-err-nm}{\vb*\eps^\nm_\gamma \coloneqq \vb Y^\nm_\gamma - \vb y^*_\gamma = B(\Sigma_\gamma)\vb*\eta_\gamma,}\end{equation}
therefore has the following approximately known distribution,
\[\vb*\eps^\nm_\gamma \sim \Norm(0, B(\Sigma_\gamma)^\top \Sigma_\gamma B(\Sigma_\gamma))\]
independently across \(\gamma\in\Gamma\). We denote the variance of each
statistic in \(\vb Y^\nm_\gamma\) by
\(\vb*\sigma^2_\gamma=\diag(B(\Sigma_\gamma)^\top \Sigma_\gamma B(\Sigma_\gamma))\).
The distribution would be exact but for discreteness in
\(\vb*\eta_\gamma\). However, for most queries and geographies, the
query variance \(\varsigma^2_{\gamma j}\) is relatively large, and so
after combining multiple queries together the discreteness artifacts are
minimal. The values of \(\varsigma^2_{\gamma j}\) are known from the
specifications of the TopDown algorithm, and we use the values provided
in the NMF.

\subparagraph{Swapping algorithm.}\label{swapping-algorithm.}

Swapping randomly moves records in the CEF from one geography to another
at a certain, but unknown, rate, which depends on a set of confidential
rules applied to the records. While opaque, swapping was applied to the
2010 census in a way that provides several benefits for end users.
First, swapping preserves the total and voting-age population of each
census block \citep{zayatz2009disclosure}. Second, most swaps occur
within small areas, so at coarser geographic scales, most data are close
to exact \citep{mckenna2018disclosure}.

Here, we represent swapping as an unknown function applied to the CEF
and a generic source of randomness \(\vb U\), which can then be
aggregated in the same was as the CEF to produce publishable statistics:
\begin{align}
    \{\vb X^\sw_\gamma\}_{\gamma\in\Gamma} &= f(\{\vb x_\gamma\}, \vb U) \\
    \vb Y^\sw_\gamma &\coloneqq A\,\vb X^\sw_\gamma
    \qfor \gamma\in\Gamma.
\end{align} We write the error produced by swapping as
\[\vb*\eps^\sw_\gamma \coloneqq \vb Y^\sw_\gamma - \vb y^*_\gamma = A(\vb X^\sw_\gamma - \vb x_\gamma).\]

\subsection{Independence Relations among the Census-induced
Errors}\label{sec-indep}

Our goal is to construct estimators of the average bias and variance of
the post-processed data \(Y^\pp_{\gamma s}\) for some set of geographies
\(\I\subseteq\Gamma\) and some statistic \(s\in\St\). We then would like
to compare these properties with those of the swapping data
\(Y^\sw_{\gamma s}\). To do so, it will be useful to state several
independence relations among the various Census-induced errors defined
above. Throughout the paper, all randomness comes from the
Bureau-injected noise \(\{\vb*\eta_\gamma\}_{\gamma\in\Gamma}\) and
\(\vb U\). The underlying CEF counts
\(\{\vb x_\gamma\}_{\gamma\in\Gamma}\) are treated as fixed. The proofs
of all results appear in Supplementary Materials Section S2.

First, some but not all of the errors are independent of each other.

\begin{prop} \label{prop-indep-methods}
\renewcommand{\theenumi}{\alph{enumi}}
For a single run of the TopDown Algorithm, the following independence relations hold:
\begin{enumerate}
\item $\{\vb*\eps^\nm_\gamma\}_{\gamma\in\Gamma}$ are mutually independent,
\item $\{\vb*\eps^\nm_\gamma\}_{\gamma\in\Gamma}$ and $\{\vb*\eps^\sw_\gamma\}_{\gamma\in\Gamma}$ are independent.
\end{enumerate}
\end{prop}

Note that \(\{\vb*\eps^\nm_\gamma\}_{\gamma\in\Gamma}\) and
\(\{\vb*\eps^\pp_\gamma\}_{\gamma\in\Gamma}\) are not independent since
the latter is a function of the former. In fact, we expect the two sets
of statistics to be positively correlated.

Furthermore, we take advantage of the fact that the Bureau released
another set of the post-processed demonstration data
\(\widetilde{Y}^\pp_{\gamma s}\) based on a different, independent, run
of the TopDown Algorithm. Although the Bureau did not release the NMF
used to produce \(\widetilde{Y}^\pp_{\gamma s}\), we know that the
released NMF, and hence \(\eps^\nm_\gamma\), is independent of
\(\widetilde{Y}^\pp_{\gamma s}\).

\begin{prop} \label{prop-indep-rerun}
\renewcommand{\theenumi}{\alph{enumi}}
For a second run of the TopDown Algorithm on the same CEF, producing errors  $\{\tilde{\vb*\eps}^\pp_\gamma\}_{\gamma\in\Gamma}$,
then 
\begin{enumerate}
\item $\{\vb*\eps^\nm_\gamma\}_{\gamma\in\Gamma}$ and $\{\tilde{\vb*\eps}^\pp_\gamma\}_{\gamma\in\Gamma}$ are independent, and
\item $\{\vb*\eps^\pp_\gamma\}_{\gamma\in\Gamma}$ and $\{\tilde{\vb*\eps}^\pp_\gamma\}_{\gamma\in\Gamma}$ are independent.
\end{enumerate}
\end{prop}

\subsection{Proposed Estimators for TopDown Bias and
Variance}\label{sec-estimators}

We wish to estimate the average bias in the TopDown statistics
\(Y^\pp_{\gamma s}\) for some set of geographies \(\I\subseteq\Gamma\)
(e.g., census tracts within a county) and some statistic \(s\) (e.g.,
the total number of voting-age Hispanics). When \(s\) is the total
population, the true CEF counts \(Y_{\gamma s}\) are known exactly
because of swapping's population invariance. Therefore,
Figure~\ref{fig-pop-boxplots} shows the exact distribution of
\(|\eps^{\pp}_{\gamma s}|\).

\textbf{Average Bias.} When \(s\) is the population of racial groups as
in Figure~\ref{fig-rmse-national}, Figure~\ref{fig-bias}, and
Figure~\ref{fig-rmse}, the true CEF counts are unknown so we must
estimate the bias from comparing the TopDown and swapping counts to the
NMF. The average bias of TopDown across these geographies may be written
as \[
\mu^\td_{\I s} = \frac{1}{|\I|}\sum_{\gamma\in\I} \E[\eps^\pp_{\gamma s}],
\] where the average bias in the post-processing is the same as the
average bias of the entire TopDown algorithm, because the NMF adds mean
zero noise and has no bias.

We propose the following simple unweighted estimator of the average
bias: \begin{align}
    \hat{\mu}^\td_{\I s} 
    \coloneqq \frac{1}{|\I|}\sum_{\gamma\in\I} \qty(Y^{\pp}_{\gamma s} - Y^\nm_{\gamma s}).
\end{align}

Subtracting the noisy statistic from the post-processed statistic
cancels the unknown true value, so that the estimator is unbiased: \[
    \E[\hat{\mu}^\td_{\I s}] 
    = \frac{1}{|\I|}\sum_{\gamma\in\I} \E[Y^{\pp}_{\gamma s} - Y^\nm_{\gamma s}]
    = \frac{1}{|\I|}\sum_{\gamma\in\I} \E[\eps^{\pp}_{\gamma s} - \eps^\nm_{\gamma s}]
    = \frac{1}{|\I|}\sum_{\gamma\in\I} \E[\eps^{\pp}_{\gamma s}]
    = \mu^\td_{\I s}.  \] Unfortunately, the variance of this estimator
is not identifiable due to the unknown correlation between
\(\eps^\pp_{\gamma s}\) and \(\eps^\pp_{\gamma^\prime s}\). Therefore,
we propose using a similar bias estimator based on an independent set of
post-processed counts \(\widetilde{Y}^\pp_{\gamma s}\) instead of
\(Y^{\pp}_{\gamma s}\): \begin{align}
\hat{\tilde\mu}^\td_{\I s} 
    \coloneqq \frac{1}{|\I|}\sum_{\gamma\in\I} \qty(\widetilde{Y}^{\pp}_{\gamma s} - Y^\nm_{\gamma s}).
\end{align}

This estimator is similarly unbiased, i.e.,
\(\E[\hat{\tilde\mu}^\td_{\I s}] = \mu^\td_{\I s}\), but it has an
important advantage over the other estimator \(\hat\mu^\td_{\I s}\)
---its variance is identifiable due to the independence with the noisy
measurements (Proposition \ref{prop-indep-rerun}(a)). The following
proposition introduces an unbiased estimator of this variance:

\begin{prop} \label{prop-pp-bias-varest}
Define the following variance estimator for the TopDown average bias estimator of $s\in\St$ across geographies $\I\subseteq\Gamma$:
\begin{align}
    \widehat V(\hat{\tilde\mu}^\td_{\I s}) 
    &\coloneqq \frac{1}{2|\I|^2}\qty(\sum_{\gamma\in\I} Y^\pp_{\gamma s} - \widetilde{Y}^\pp_{\gamma s})^2 +
    \frac{1}{|\I|^2}\sum_{\gamma\in\I} \sigma^2_{\gamma s}.
\end{align}
Then $\widehat V(\hat\mu^{\td}_{\I s})$ is unbiased: \[
    \E[\widehat V(\hat{\tilde\mu}^\td_{\I s})] = \Var[\hat{\tilde\mu}^\td_{\I s}].
\]
\end{prop}

\textbf{Average Variance.} In addition to the average bias in the
TopDown counts \(Y^\pp_{\gamma s}\), it is also possible to obtain an
unbiased estimate of their average variance for some set of geographies
\(\I\subseteq\Gamma\) and some statistic \(s\). Unlike with the bias
estimation, however, we can no longer form an unbiased estimate with the
TopDown counts \(Y^\pp_{\gamma s}\) alone. Instead, we must use
\(\widetilde{Y}^{\pp}_{\gamma s}\), which are independent of
\(Y^\nm_{\gamma s}\). Our proposed estimator is: \[
    \widehat{V}^{\td}_{\I s} = \frac{1}{2|\I|} 
    \sum_{\gamma\in\I} \qty(Y^\pp_{\gamma s} - \widetilde{Y}^{\pp}_{\gamma s})^2.
\] This estimator is unbiased for
\(V^\td_{\I s}\coloneqq|\I|^{-1}\sum_{\gamma\in\I}\Var[\eps^\pp_{\gamma s}]\)
according to the proof of Proposition \ref{prop-pp-bias-varest}. We use
this estimator to compute approximate confidence intervals in
Figure~\ref{fig-bias}. Note that the confidence intervals for the total
population column in Figure~\ref{fig-bias} is estimated differently,
because the unit-level error of those from a draw of the TopDown
algorithm are known exactly. We have two TopDown draws, and each
provides one estimate. For each population bin, we use the average of
these two draws as our point estimate, and compute its standard error as
\(\widehat V(\hat\mu^{\pp}) =  \frac{1}{2} |Y^{\pp} - Y^\nm|\).

\textbf{Mean Square Error.} Finally, we can obtain an unbiased estimate
of the mean square error (MSE) for TopDown counts, which is defined as
\[ 
\mse^\td_{\I s} = \frac{1}{|\I|}\sum_{\gamma\in\I} \E[{\eps^\pp_{\gamma s}}^2]\]
and represents the combination of bias squared and variance. We propose
the following estimator \[
    \widehat{\mse}^{\td}_{\I s} 
    \coloneqq \frac{1}{|\I|}\sum_{\gamma\in\I} 
    \left[(\widetilde{Y}^{\pp}_{\gamma s}-Y^\nm_{\gamma s})^2 - \sigma^2_{\gamma s}\right],
\] which is unbiased for the MSE: \[
    \E[\widehat{\mse}^{\td}_{\I s}]
    = \frac{1}{|\I|}\sum_{\gamma\in\I} 
    \E\qty[(\widetilde{Y}^{\pp}_{\gamma s}-Y^\nm_{\gamma s})^2] - \sigma^2_{\gamma s}
    = \frac{1}{|\I|}\sum_{\gamma\in\I} 
    \E[(\tilde{\eps}^\pp_{\gamma s})^2] + \sigma^2_{\gamma s} - \sigma^2_{\gamma s}
    = \mse^\td_{\I s}.\]

In practice, this estimator may take a negative value due to a high
degree of estimation noise. In such cases, we replace a negative
estimate with zero in our analysis. This occurs in
4 out of the 40
\unskip ~TopDown estimates in
Figure~\ref{fig-rmse-national}, but occurs in none of the TopDown
estimates in Figure~\ref{fig-rmse}. We take the square root of this
estimator to obtain the estimated root mean square error (RMSE) in
Figure~\ref{fig-rmse-national} and Figure~\ref{fig-rmse}.

\subsection{Proposed Estimators for Swapping Bias and
Variance}\label{proposed-estimators-for-swapping-bias-and-variance}

Estimating the average bias of the swapped counts \(Y^\sw_{\gamma s}\)
is also straightforward, and we can simply apply the same strategy as
the one used for the TopDown algorithm. \[
    \E[\hat{\mu}^\sw_{\I s}] 
    = \frac{1}{|\I|}\sum_{\gamma\in\I} \E[Y^{\sw}_{\gamma s} - Y^\nm_{\gamma s}]
    = \frac{1}{|\I|}\sum_{\gamma\in\I} \E[\eps^{\sw}_{\gamma s} - \eps^\nm_{\gamma s}]
    = \frac{1}{|\I|}\sum_{\gamma\in\I} \E[\eps^{\sw}_{\gamma s}]
    = \mu^\sw_{\I s}.  \] Unfortunately, the valid standard error is
difficult to obtain because, unlike the case of the TopDown algorithm,
we do not have access to an independent second draw. To address this
unidentifiability, we propose the following estimator, which is
conservative under an additional assumption.

\begin{prop} \label{prop-sw-bias-varest}
Define the following variance estimator for the swapping average bias estimator of $s\in\St$ across geographies $\I\subseteq\Gamma$:
\begin{align}
    \widehat V(\hat\mu^{\sw}_{\I s}) 
    &\coloneqq \frac{1}{|\I|^2}\sum_{\gamma\in\I} (Y^\sw_{\gamma s} - Y^\nm_{\gamma s})^2 \\
\end{align}
If \(
    \sum_{\gamma\neq\gamma^\prime\in\I} \Cov(\eps^\sw_{\gamma s}, \eps^\sw_{\gamma^\prime s}) \le 0,
\) then \[
    \E[\widehat V(\hat\mu^{\sw}_{\I s})]\ge \Var[\hat\mu^{\sw}_{\I s}]. 
\] 
\end{prop}

The assumed weakly negative correlation across geographies is plausible,
since errors in one geography must be matched by opposite errors in
other geographies in order for the error at the higher geographic level
to be controlled. The lack of detailed information about the swapping
algorithm, however, makes it difficult to know whether this additional
assumption holds.

We can estimate the MSE of the swapped counts \(Y^\sw_{\gamma s}\)
similarly as with the TopDown counts, using the following estimator \[
    \widehat{\mse}^{\sw}_{\I s} 
    \coloneqq \frac{1}{|\I|}\sum_{\gamma\in\I} 
    (Y^\sw_{\gamma s}-Y^\nm_{\gamma s})^2 - \sigma^2_{\gamma s},
\] which is unbiased for \(\mse^\sw_{\I s}\) by an identical argument to
that for the unbiasedness of \(\widehat{\mse}^\pp_{\I s}\).
Unfortunately, like the case of the TopDown algorithm, this estimate may
take a negative value in practice. This occurs in
3 out of the 40
\unskip ~Swapping estimates in
Figure~\ref{fig-rmse-national}, but occurs in none of the TopDown
estimates in Figure~\ref{fig-rmse}. Again, a negative value will be
replaced with zero in our analysis.

\section{Results}\label{sec-results}

\subsection{Absolute errors in total population counts are small on
average}\label{sec-abserror}

We first examine errors in the total population counts in different
census geographies. To do so, we compare counts from the post-processed
and noisy measurement files against the counts from the swapping
algorithm in the original 2010 decennial Census release. The Bureau's
particular swapping algorithm preserved a geographic unit's total
population from the CEF, because it swapped households with the ``same
number of adults {[}and{]} same number of children''
\citep[p.~518]{censusSF1docs}. Therefore, counts for total population
released after swapping serves as the ground truth against which we
measure TopDown algorithm's error exactly \citep{kenny2023comment}. No
statistical assumptions are necessary to compute this error.

\begin{table}

\caption{\label{tab:sizes}\textbf{Population summary statistics for Census geographies.}
               Summaries across populated geographic units studied in this paper.
               Blocks are nested in block groups, which are nested in tracts.
               Place stands for Census Designated Place.
               For example, the median Census Block is populated by 23 people.
               }
\centering
\begin{tabular}[t]{lccccc}
\toprule
  & Block & Block Group & Voting District (VTD) & Tract & Place\\
\midrule
Minimum & 1 & 1 & 1 & 1 & 1\\
25th percentile & 8 & 912 & 734 & 2,901 & 331\\
Median & 23 & 1,242 & 1,215 & 4,014 & 1,068\\
75th percentile & 54 & 1,726 & 2,084 & 5,333 & 4,038\\
Maximum & 19,352 & 37,452 & 53,877 & 37,452 & 8,175,133\\
\bottomrule
\end{tabular}
\end{table}

We reconstructed the NMF to create a national dataset for five census
geographies: blocks, (tabulation) block groups, tracts, voting
districts, and census places (see Materials and Methods,
Section~\ref{sec-addl}). Total population quantiles for these
geographies among populated geographic units are reported in Table
\ref{tab:sizes}. Blocks, block groups, and tracts are statistical units
used by the Census Bureau, while voting districts generally represent
voting precincts and census places typically represent towns or cities.

A challenge in comparing these counts to the NMF is that geographic
hierarchies, or \emph{spines} in Census terminology, are defined
differently in the NMF than in the standard Census data (see
Figure~\ref{fig-nmf-heirarchy}). We therefore parsed the NMF geography
identifiers and transformed them into their regular Census spine
equivalents. This results in several modifications, detailed in
Materials and Methods, Section~\ref{sec-addl}. For example, the NMF
treats portions of a tract that are designated as \aian{} (American
Indian / Alaska Native) separately from the rest of the tract by
assigning it different degrees of privacy-protecting noise. In these
cases we combine the estimates of both the \aian{} and non-\aian{}
fragments to a single tract-level NMF estimate. Thus, the estimates we
use match the geographies for which traditional census data is released.

When multiple estimates are produced for each geography in the NMF, we
combine them by following the method of \citet{mccartan2023nmf}. We
expect that \emph{on-spine} geographies, i.e.~geographies part of the
NMF spine (blocks, tracts, and places), will suffer from less bias and
noise than \emph{off-spine} geographies (non-optimized block groups and
voting districts) that can receive more noise injection by virtue of
containing fragments of on-spine geographies.

\begin{figure}[tb]

\centering{

\includegraphics[width=1\textwidth,height=\textheight]{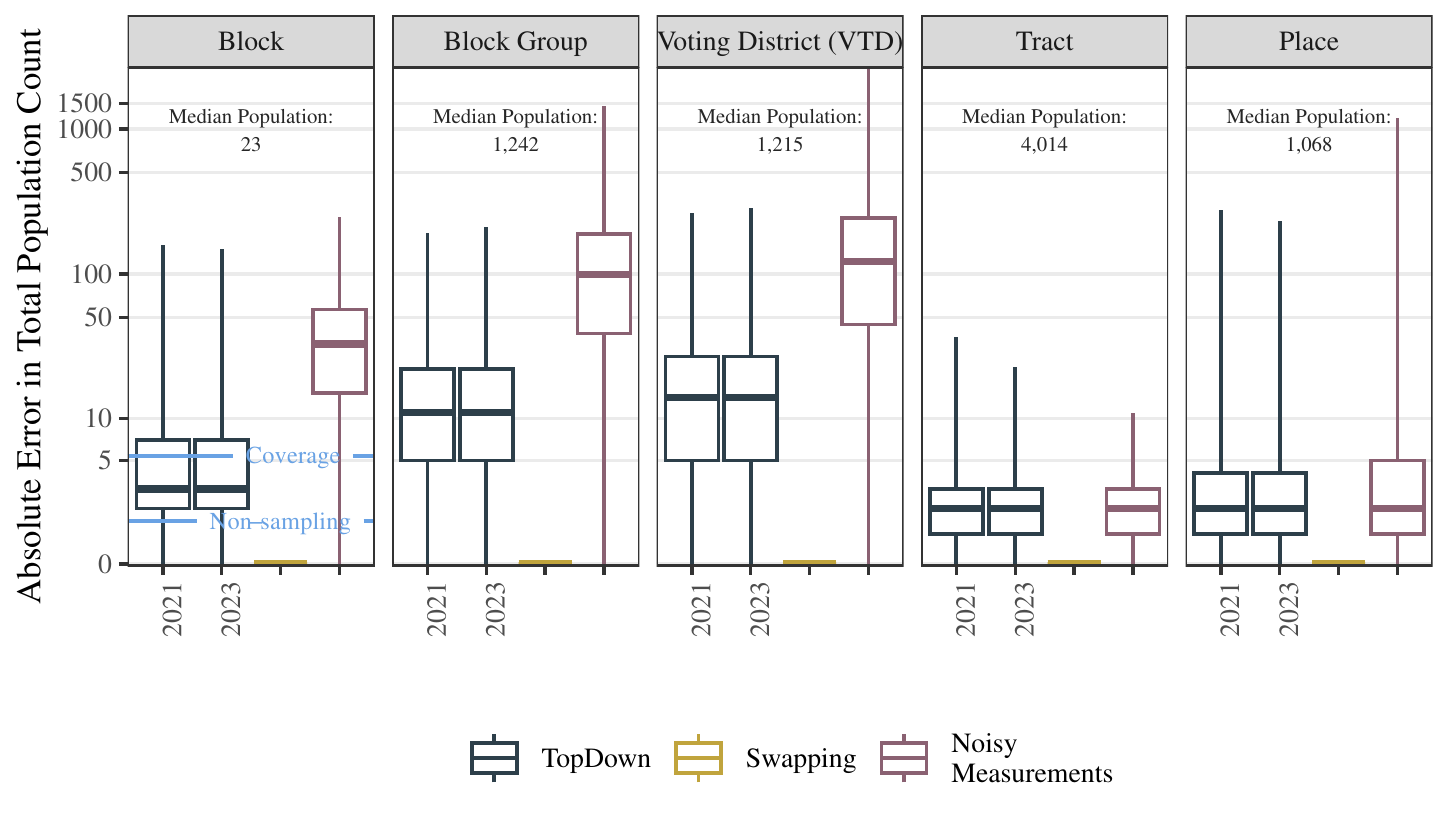}

}

\caption{\label{fig-pop-boxplots}\textbf{Distribution of absolute count
error in total population.} The figure displays the absolute error in
total populations as enumerated in the Census Edited File (CEF). The
\emph{y}-axis is shown on a pseudo-log10 scale. Each panel depicts a
geographical level where a boxplot shows the nationwide distribution of
population error at that geographical level, with horizontal lines for
the first quartile, median, and third quartile. Swapping errors are
always zero for census geographies, as total population remains
invariant. The Bureau-estimated mean absolute error due to coverage and
non-sampling errors at the block level are indicated on the leftmost
pattern by horizontal lines. Coverage and non-sampling error affect the
CEF and thus are present in addition to DAS errors from TopDown,
swapping, or noisy measurements.}

\end{figure}%

Figure~\ref{fig-pop-boxplots} shows the absolute error in counts, that
is, the absolute difference between a geographic unit's total population
from each algorithm and the corresponding count in the CEF. We find
consistently that the NMF has substantially larger errors than TopDown
for both small and off-spine geographies. These NMF errors shrink to
become comparable to the TopDown errors for larger geographies. The
estimated absolute TopDown error for blocks is typically three to seven
people after post-processing, though there is a substantial tail of
larger errors. Noisy measurements often have much larger errors,
frequently exceeding the size of the median populated block, which is
only 23 people. Thus in relative terms, the TopDown errors are on the
order of 10\% of the typical block's population. Swapping, as noted
above, preserves the total population of every census geography as an
invariant and so its error is exactly zero for all census geographies.
Note that any geography built from whole blocks, which includes all
census geographies, will then have zero total population error from
swapping (For the hierarchy of census geographies which can be built
from whole blocks, see
\url{https://www2.census.gov/geo/pdfs/reference/geodiagram.pdf}).

Block groups and voting districts (VTD) exhibit similar patterns, where
the TopDown errors are smaller than the noisy measurement errors.
However, tracts show nearly-equivalent errors across both TopDown and
noisy measurements. Notably, tracts are on-spine geographies, so each
tract receives dedicated queries in the noising process. Geographies
that are small and off-spine, like block groups and VTDs, have larger
errors in the NMF. Absolute errors are typically very small for census
places, as displayed in the rightmost panel of
Figure~\ref{fig-pop-boxplots}. Census places vary greatly---from small
incorporated places to New York City---and we explore possible
heterogeneity across population size in Section~\ref{sec-deciles}.

In sum, the typical absolute error in total population is small in
post-processed data, while the noisy measurement errors are much
greater. These errors are largest in relative terms for small places,
especially those which are off-spine geographies. Errors are likely
larger for off-spine geographies because they do not benefit from the
same efforts to minimize errors that TopDown incorporates for on-spine
geographies. Notably, many other geographies used in practice will also
be off-spine, including legislative districts, school districts,
neighborhoods, and individual Native American reservations.

These estimates of DAS bias are roughly similar in magnitude to the two
main sources of non-DAS error---coverage error (omissions and incorrect
enumerations) and non-sampling error (such as data entry errors or
incorrect locations) at the block level. Non-sampling errors can come
from sources like transcription errors and incorrectly locating housing
units. Coverage error results from omissions or additional incorrect
enumerations; i.e., differences between the target population and
population actually counted by the Census. The Bureau has estimated
approximate sizes of these errors in originally internal memos at the
block and county levels \citep{memo21}. At the block level, the mean
absolute coverage error is approximately 5.5 people; the mean absolute
non-sampling error is 1.5 people \citep[Table 2 pg. 23]{memo21}. In
comparison, the block-level mean absolute error for TopDown is 4.9 and
for the noisy measurements is 39.1. This means that the block-level
variance in population counts is increased by about 75\% due to the new
TopDown algorithm (since the TopDown and coverage errors are
statistically independent, the new variance is
\(1.5^2 + 5.5^2 + 4.9^2\approx 56.5\), versus the non-DAS error variance
of \(1.5^2 + 5.5^2\approx 32.5\)---a roughly 75\% increase).

The share of the total census error which is attributable to privacy
protections is highest at the block level, and will decrease for larger
geographies. Note that the Bureau has not reported the results of their
error simulations for geographies other than counties and census blocks.
However, the magnitude of coverage and non-sampling error would increase
with population. At the same time, larger Census geographies have
smaller DAS error because of the design of the TopDown algorithm.

While the noise added to tabulations to produce the NMF is completely
random, we note that the TopDown errors shown in
Figure~\ref{fig-pop-boxplots} are non-random because of post-processing.
The error produced by one run of the TopDown algorithm for a given
geography will typically be similar to the error produced for that same
geography on a completely independent run of the TopDown algorithm. This
is due to the post-processing step, which incorporates the complex
interplay of population statistics, the structure of the geographic
spine, and the constraints and invariants imposed by the Bureau. Table
\ref{tab:corr} reports the sample correlation in the population error
between the two independent runs of the TopDown algorithm (the 2021 and
2023 demonstration data), across geographies, for a given level. Even
for typically populous geographic units like Census places, the sample
correlation in errors across geographies can be as high as 0.63.
Roughly, this means that around \(0.63^2=40\%\) of the variance
introduced by the TopDown algorithm is systematic rather than random.
This does not appear to be an artifact of the necessary upward bias for
geographies with zero population: the pattern remains consistent even
among geographies with at least 100 or 1,000 people.

\begin{table}

\caption{\label{tab:corr}\textbf{Correlation of TopDown errors between runs}.
               This table reports the sample correlation between the TopDown 
               errors in the 2021 and 2023 demonstration data across each
               kind of Census geography. Correlations are calculated for
               errors in total and voting-age population because the exact
               counts for those statistics are invariant under swapping.}
\centering
\begin{tabular}[t]{llllll}
\toprule
  & Block & Block Group & Voting District (VTD) & Tract & Place\\
\midrule
Voting-age population & 0.280 & 0.409 & 0.393 & 0.104 & 0.634\\
Total population & 0.449 & 0.619 & 0.600 & -0.007 & 0.572\\
\bottomrule
\end{tabular}
\end{table}

\subsection{Counts of racial and ethnic groups are accurate under
TopDown and swapping}\label{sec-deciles}

Unlike the total population counts, we do not have access to the ground
truth counts for each racial and ethnic group because they are not held
invariant during swapping. We thus must construct estimators for the
\emph{average} bias and root mean square error of both TopDown and
swapping, using the unbiased but noisy NMF. The validity of our
estimators only relies on the fact that swapping, one run of the NMF and
TopDown algorithm, and another run of the NMF and TopDown algorithm are
all based on entirely independent sources of randomness. This
proposition, formalized in Materials and Methods
Section~\ref{sec-estimators}, is justified by official documentation by
the Bureau.

The root mean square error (RMSE) of TopDown counts represents the
square root of the combination of the bias (squared) and variance of
TopDown. The bias originates from the post-processing, whereas the
variance originates from both the NMF and the post processing. Materials
and Methods Section~\ref{sec-estimators} shows how we leverage the NMF
to construct an unbiased estimator for the mean square error (MSE).

\begin{figure}[t]

\centering{

\includegraphics[width=1\textwidth,height=\textheight]{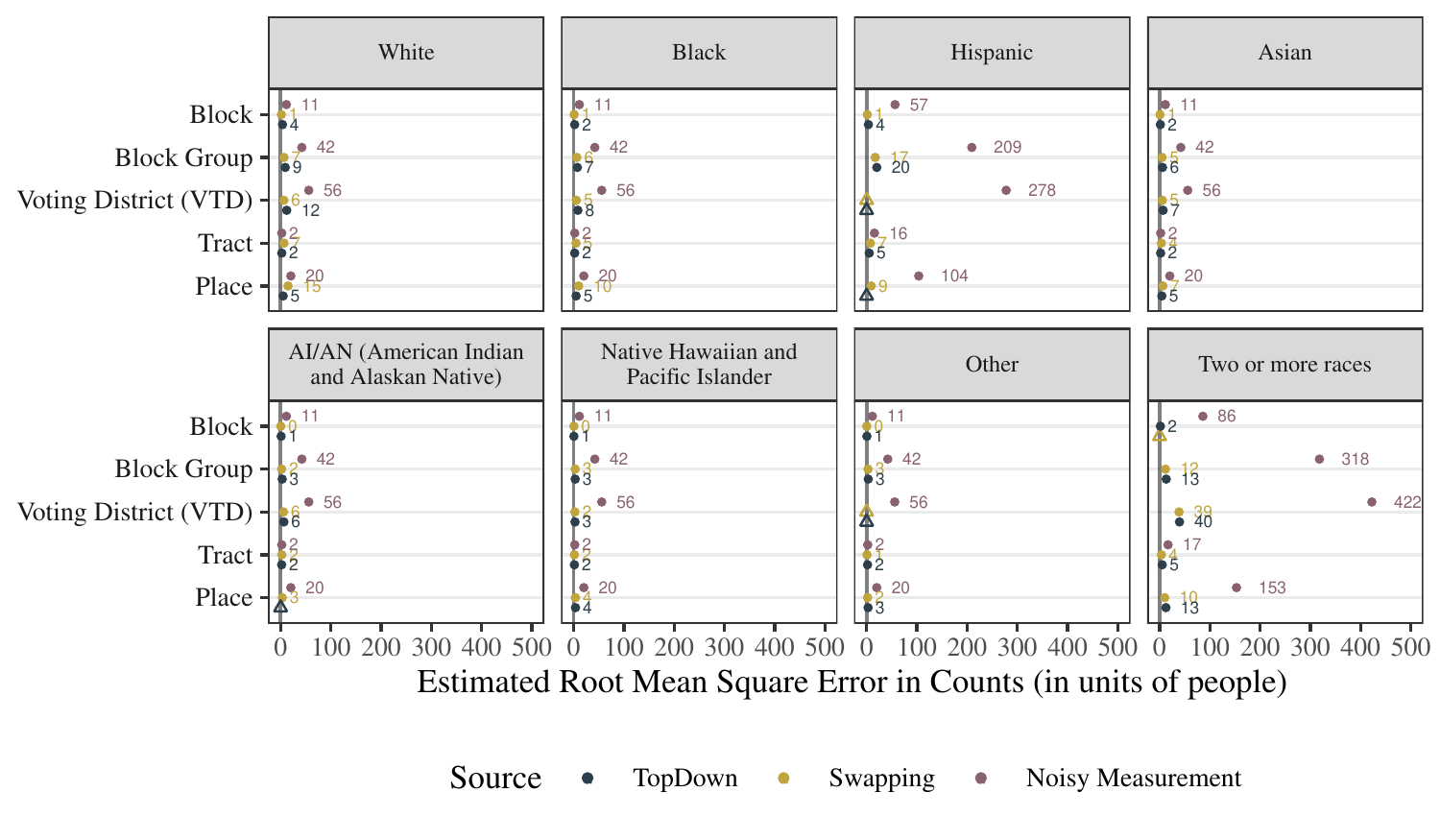}

}

\caption{\label{fig-rmse-national}\textbf{Estimated root mean square
error (RMSE) for population counts of a race/ethnicity group, at each
geographic level.} The RMSE quantifies the average magnitude of error
for a given geography for a particular geographic unit (see
Section~\ref{sec-estimators} for estimators). Triangles for RMSE
indicate that the estimated mean square error was negative and hence was
set to zero.}

\end{figure}%

Figure~\ref{fig-rmse-national} demonstrates that the estimated RMSEs for
TopDown and swapping are small in absolute terms for most race and
ethnicity groups across census geographic levels. We find little
evidence that TopDown and swapping are substantially different in terms
of RMSE. We do, however, caution that the RMSE estimates for TopDown and
swapping are particularly noisy due to the high variability of the noisy
measurements, and lack uncertainty quantification that is available for
our average bias estimates.

On the other hand, because the variance of the NMF counts is reported by
the Bureau and known exactly, the RMSE of the NMF can be calculated
exactly. The results show that the NMF counts are much noisier than the
post-processed TopDown and swapping counts. The errors are particularly
substantial for off-spine geographies such as block groups, VTDs, and
places as well as for Hispanics, multi-racial groups. This is because
these groups are a combination of multiple queries
\citep{mccartan2023nmf}. In the Census classification, Hispanic voters
consist of Hispanic Whites, Hispanic Blacks, and other racial groups,
whereas non-Hispanic Whites consist only of one racial group. Computing
the Hispanic population in an off-spine geography would therefore
involve identifying a set of on-spine geographies that consist of the
target off-spine geography and summing the population counts of
subgroups that form Hispanics. This summation process across racial
categories and geographies leads to a greater variance of the resulting
population count (within each on-spine geography, each race query is
given the same variance). Post-processing of the NMF reduces this
variance substantially.

Next, we turn to separating out the bias component of the RMSE. The
\emph{average} bias represents the average of biases across geographic
units. Since each bias could be either positive or negative, the average
bias can mask important variation of biases that may exist in some
geographies. For example, although the average of biases across all
geographic units in the nation is necessarily zero, that does not mean
the bias is negligible for any given unit.

Indeed, the average bias is likely much smaller than the absolute
average bias shown in Figure~\ref{fig-pop-boxplots}. We can compute the
average bias by leveraging the NMF as a comparison. In addition, we can
compute the estimation uncertainty around our average bias estimate by
leveraging the fortuitous existence of two draws of the TopDown
algorithm (Materials and Methods Section~\ref{sec-estimators}).

To test whether the bias tends to be positive in some geographies but
not others, we classify geographic units into ten bins of total
population, and estimate the average bias of a racial group count within
that subset of geographic units. Accurate binning of this sort is
possible because the total population counts are reported without
privacy-induced error under swapping.

\begin{figure}[t]

\centering{

\includegraphics[width=1\textwidth,height=\textheight]{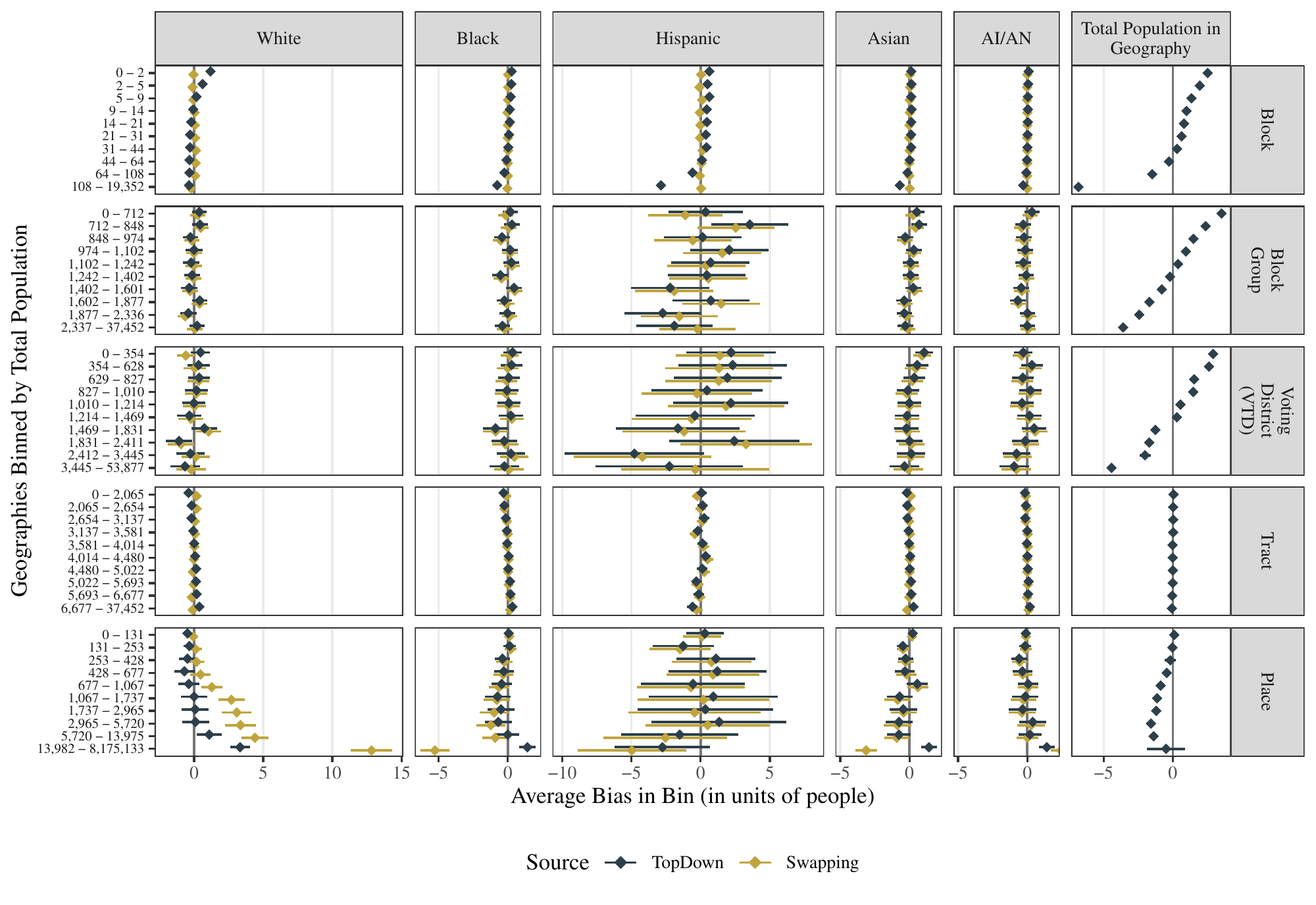}

}

\caption{\label{fig-bias}\textbf{Average bias for race/ethnicity
population counts at each geographic level by its total population.} The
figure estimates the average overcounting or undercounting in a group of
geographies, separately for five geographic levels and five
race/ethnicity groups. See Section~\ref{sec-estimators} for estimators.
Bins on the \emph{y}-axis are deciles of total population of the
geographic level Points show the estimated bias, and lines show
estimated 95\% confidence intervals.}

\end{figure}%

The result, Figure~\ref{fig-bias}, shows that estimated average biases
are consistently small across all racial groups, geographic levels, and
data sources. Figure S1 in the Supplementary Materials demonstrates
these patterns for the race groups that are not shown in
Figure~\ref{fig-bias}. For on-spine geographies (block and tract),
average biases are estimated to be consistently near zero in almost all
cases. Off-spine geographies (block group, VTD, and place) have
estimated average biases with slightly higher magnitudes, but they are
rarely statistically distinguishable from zero. These small average
biases appear to address patterns found during previous releases of DAS
demonstration data (further details on the updates made to the DAS after
the April 2021 Demonstration Data are available at
\url{https://www.census.gov/programs-surveys/decennial-census/decade/2020/planning-management/process/disclosure-avoidance/2020-das-updates/2021-06-09.html}).
The improvements could be due to increases in the privacy budget or
changes to the TopDown algorithm.

Although the magnitude of bias is not large, small racial counts are
biased systematically. In the largest Census Places, swapping tends to
overcount White people, undercount Hispanic, Black, and Asian people.
TopDown has smaller average bias in these geographies.

TopDown does induce a detectable, though again small, systematic pattern
of counting bias for total population. It tends to overcount total
populations in small blocks, block groups, and VTDs by about 2-3 people,
and undercount them in large blocks, block groups, and VTDs. We computed
average biases for total population using the average of two TopDown
draws compared with the Swapping algorithm, and without relying on the
NMF.

\begin{figure}[t]

\centering{

\includegraphics[width=1\textwidth,height=\textheight]{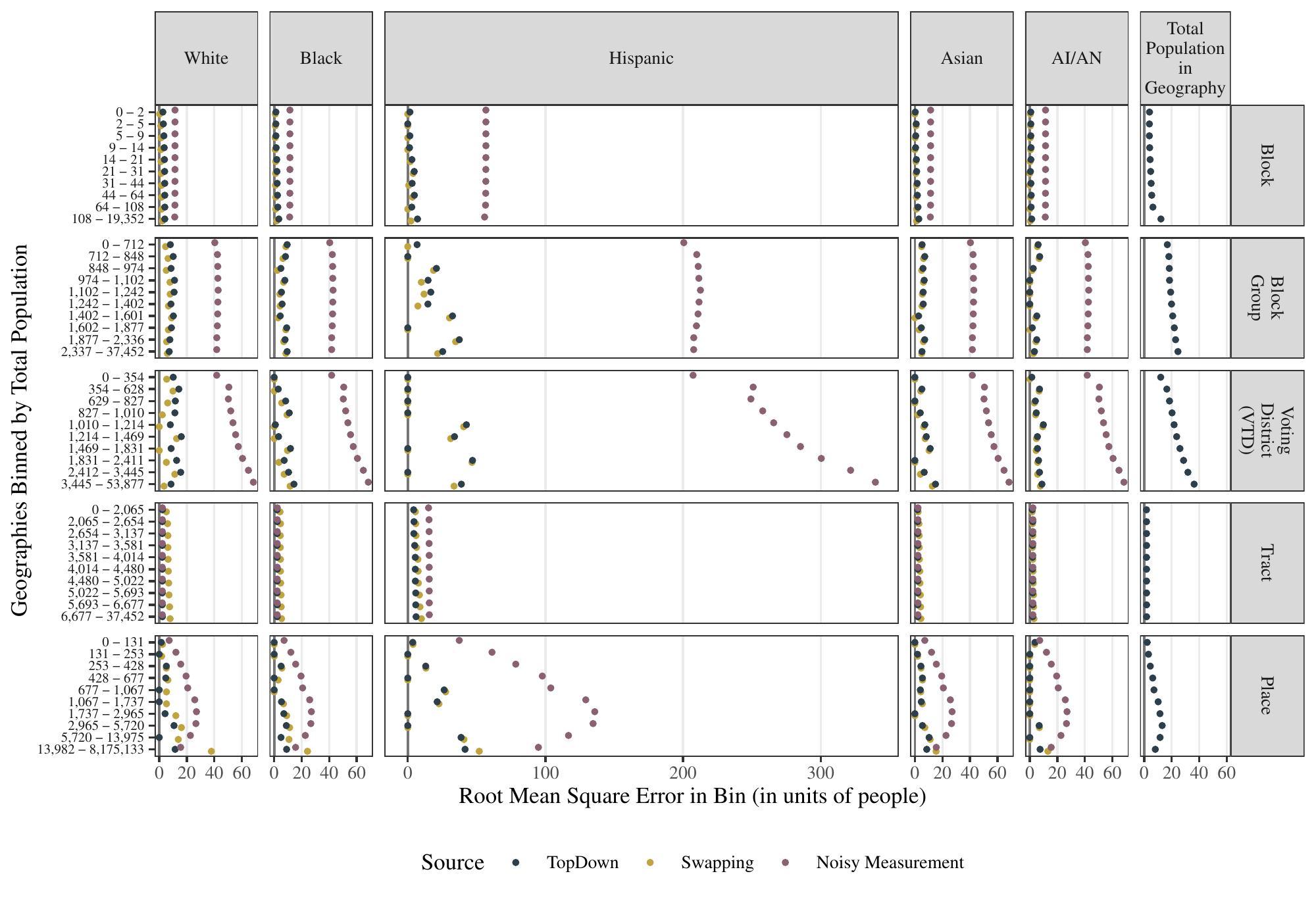}

}

\caption{\label{fig-rmse}\textbf{Estimated root mean square error of
race/ethnicity counts at each geographic level, by its total
population.} Unlike Figure~\ref{fig-rmse-national}, it estimates RMSE
for a subset of geographies. See Section~\ref{sec-estimators} for
estimators. Bins on the \emph{y}-axis are deciles by total population of
the geographic level.}

\end{figure}%

We also calculate RMSE by population size bin (Figure~\ref{fig-rmse} and
Figure S3 in the Supplementary Materials). As expected, we find largely
constant RMSE for NMF across on-spine geographies (block and tracts).
RMSEs for off-spine geographies (block group, VTD, and place) generally
stay constant or increase with the size of the population decile. As in
Figure~\ref{fig-rmse-national}, we find consistently larger RMSEs for
the Hispanic population.

NMF consistently has the largest errors, but these errors are
significantly reduced by post-processing. While our estimates are
necessarily noisy given that the Bureau has only released two
independent runs of the TopDown algorithm under the final set of
parameters, we find no evidence of large systematic errors. Instead, our
findings suggest that post-processing significantly improves the data
accuracy. The resulting TopDown estimates are roughly as variable as
those under swapping in nearly all instances. Of course, the
post-processing introduces an intractable dependency structure to the
errors, while the NMF errors are independent and have a known
distribution. Thus despite the higher variance, the NMF counts can be
used to produce statistically valid downstream estimates.

Further, these small bias and uncertainty estimates are largely constant
across population size. We stress that this is unlike other sources of
Census error, which Census researchers have found tend to scale with
population size \citep{memo21}. For example, Census researchers estimate
that coverage error in blocks with populations over 1,000 have mean
absolute deviations dozens of times larger than blocks with 1-9 people
(14.5 vs.~0.4) \citep[Table 5 pg. 15]{memo21}. In contrast, we estimate
that errors from TopDown, NMF, and swapping are approximately constant
across all population deciles, as shown in Figure~\ref{fig-rmse}. While
errors from NMF tend to be larger than TopDown, swapping, and other
sources of error for small geographies, error sizes are roughly constant
across population size for on-spine geographies. We find NMF errors in
off-spine geographies can broadly increase across population sizes, but
even these are generally reduced to near zero for TopDown and swapping
estimates.

These findings imply that for large-population geographies at all
levels, variation from the TopDown and swapping algorithms is likely
much smaller than variation from coverage and non-sampling errors.
However, for small-population geographies, where coverage and
non-sampling errors are much smaller, variation from the TopDown and
swapping algorithms can be much greater relative to both the census
errors and the size of the population. For example, the typical RMSE for
race counts at the block level is about 3 people. Since the median
populated block has only 23 people, this means that the variation in
race counts is on the same order of magnitude as the actual population
of each race. And for smaller racial groups, the RMSE can be larger than
the population of the racial group itself.

Census data users should thus take note: for small-population
geographies, the size of the DAS-introduced error can be significant
relative to the true CEF counts of interest as well as other census
errors. This warning applies not only for the new TopDown algorithm but
also for the swapping algorithm used in the three previous decennial
censuses.

\subsection{Accuracy does not substantially vary with racial
heterogeneity}\label{accuracy-does-not-substantially-vary-with-racial-heterogeneity}

Finally, we assess whether bias in total population varies by the
demographics of the geography, instead of the size of the geography. We
focus on the percentage of the geography that identifies as non-White.
We use this proportion, rather than a direct measure of diversity like
the Herfindahl-index, to avoid transformations that may introduce bias.
This relationship is difficult to assess because racial group counts
(unlike total population) themselves are measured with swapping or
noisy-measurement error. The above analysis has shown that the average
magnitude of the noisy-measurement error is an order of magnitude larger
than the average noise introduced by the TopDown algorithm.

To overcome this challenge, we fit a flexible nonparametric regression
model that accounts for the known noisy measurement error structure in
racial group counts using the method of \citet{fan2009JASA} as
implemented by \citet{lpme}. This allows us to properly estimate the
average bias in each geography (measured without error) as a function of
the percentage of the population that is non-White (which is measured
with error). The details about implementation are described in
Supplementary Materials Section S4. We note that the original method of
\citet{fan2009JASA} does not allow for the between-geography correlation
that is present here. However, we expect this correlation to
significantly affect the variance, rather than the bias estimates we
present. While this method does not provide measures of uncertainty,
simple binned averages using the methodology of Section~\ref{sec-addl}
also yields the same pattern, where undercounting biases in block groups
and voting tabulation districts are statistically distinguishable (see
Figure S3 in the Supplementary Materials). This approach ignores the
noise in the percentage non-White residents, but the noise is on the
order of a few percentage points, and if anything would be expected to
attenuate the observed pattern, not contribute to it.

\begin{figure}[t]

\centering{

\includegraphics[width=1\textwidth,height=\textheight]{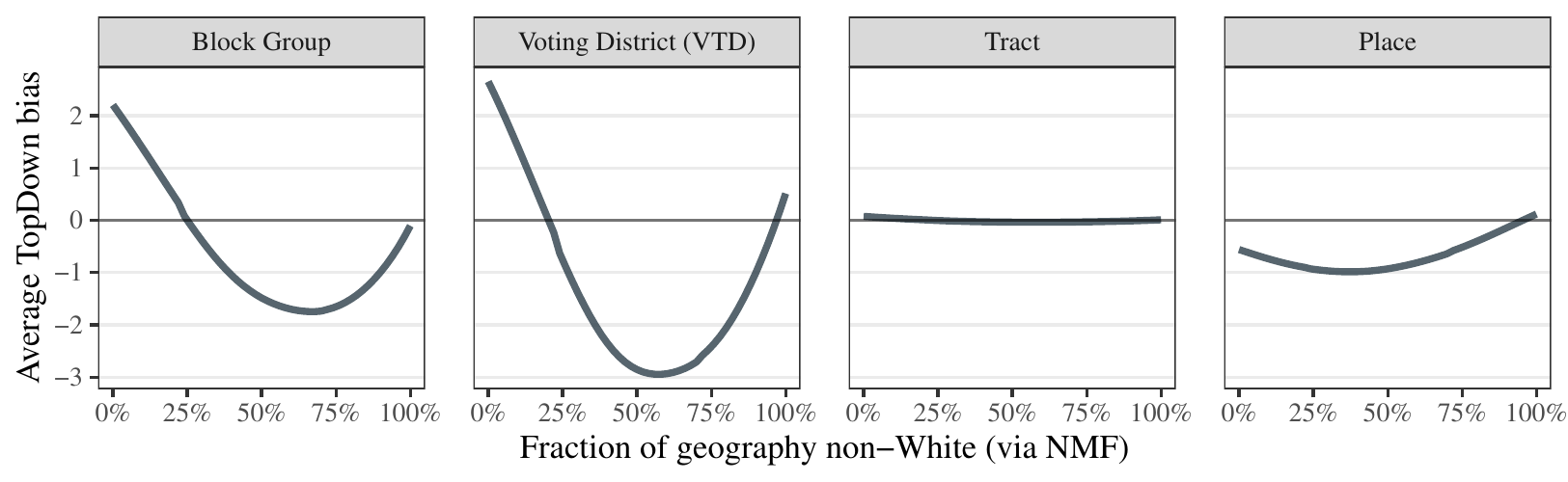}

}

\caption{\label{fig-lpme-main}\textbf{Average bias in total population
count by percent non-white.} The figure estimates the degree of average
TopDown overcounting or undercounting by the racial demographics of the
geographic unit. Lines are estimated from a local polynomial regression
model that adjusts for noise in the non-white percentage estimate.
Swapping would have no error for total population.}

\end{figure}%

Figure~\ref{fig-lpme-main} shows the fitted local polynomial regression
lines for four geographies. For tracts, which is an on-spine geography,
the pattern of average bias is essentially flat with no bias. Off-spine
geographies exhibit a systematic relationship where geographies that are
30 to 70 percent non-White are systematically undercounted (consistent
with \citet{kenny2021impact}). However, the magnitude of average
undercounting is only around 1 to 3 people, consistent with earlier
results. Recall that the swapping algorithm preserves total population
and hence has no error for any census geographical unit.

The modest biases may be due to various changes the Bureau made to the
DAS post-processing procedure and privacy budget throughout feedback
received during several ``demonstration data'' releases. For example,
\citet{censusparams} documents that the DAS team made changes to the
post-processing system to reduce biases undercounting population in
racially and ethnically diverse areas, as documented in
\citet{kenny2021impact}. Earlier work in \citet{kenny2021impact} found
that population biases likely decreased between iterative demonstration
releases of the DAS after the Bureau incorporated several changes from
stakeholder feedback.

\section{Discussion}\label{sec-discussion}

We have shown how to use the recently released noisy measurement file
(NMF) to evaluate the bias and noise caused by the Census Bureau's
disclosure avoidance systems (DAS). The primary advantage of our
methodology is the ability to characterize the DAS-induced errors
relative to the Census Edited File (CEF), which is the non-privatized
confidential census microdata. Although the CEF remains private, we are
able to evaluate bias and noise by exploiting known statistical
properties of the noise induced to create the NMF.

Previous studies, without access to the NMF, were only able to compare
the data sets processed by two different DAS; TopDown and swapping
algorithms. These past comparisons remain useful for policy evaluation
and allow researchers to compare how policy outcomes may change when
basing decisions on a new dataset. However, they could not assess the
relative magnitude of bias and noise induced by the TopDown and swapping
algorithms because their analyses depend on comparisons between versions
of the DAS \citep{boydSarathy, kenny2023comment}.

Our evaluation has shown that the average biases introduced by both the
TopDown and swapping algorithms are overall small. The variances of the
TopDown and swapping error are also similar in magnitude. These
DAS-induced errors are usually smaller than other sources of census
error, and also tend to be small relative to the population of each
geography. These findings can reassure users of Census data that
population counts from large-population geographies, like counties,
generally include only minor errors induced by both swapping and
TopDown.

For small-population geographies like, however, both TopDown and
swapping algorithms introduce larger relative errors. We emphasize two
points about counts from small-population areas for Census data users to
keep in mind. First, estimates from small-population geographies like
census blocks have the largest variance among census geographies. We
find these deviations can be as large or larger than other sources of
census error like coverage and non-sampling error. Second, this variance
at small levels can be large relative to the geographies' populations.
Consequently, analysts of Census data since 1990 should proceed
carefully in analyzing small areas and counts, since they may contain
more relative noise from the DAS than previously understood.

The errors in the NMF are substantially larger than that from TopDown
and swapping, but they come with statistical guarantees that enable
researchers to conduct valid statistical inference relative to the
confidential CEF. Indeed, the estimates presented here were only made
possible by these statistical properties of the NMF. Thus, while
analyses that treat the NMF counts as ground truth (i.e., as a drop-in
replacement for the published Census statistics) will likely result in
much higher bias and variance, the NMF should play an important role for
researchers working with Census data.

These results have several policy implications for future releases of
decennial Census data. First, post-processing plays an important role in
preserving the accuracy of Census data for direct use. The NMF, while
unbiased, may not be sufficiently accurate for public policy decisions
without measurement error modeling and using it naively could have
differential impacts on racial and ethnic groups. The final small biases
and noise come after rounds of iteration between the Bureau and data
stakeholders, which our results here and elsewhere suggest likely
improved the accuracy of the final released data
\citep{kenny2021impact}.

Second, while the Bureau replaced swapping with TopDown, citing the need
for formal privacy guarantees to address increasing vulnerability to
privacy leaks, we find that the bias and noise induced by two algorithms
are similar. Our finding is consistent with our earlier result that an
individual's race can be predicted accurately under both swapping and
TopDown \citep{kenny2021impact}. It is also consistent with the recent
analysis that shows swapping can be differentially private, albeit with
a different differential privacy definition \citep{bailie2023swap}.

Taken together, these results provide additional empirical context for
ongoing conversations about the Bureau's trade-off between data accuracy
and privacy protection. Biases from data produced by the TopDown and
swapping algorithms are smaller than found in the released NMF, but the
extent to which privacy is meaningfully protected is disputed and
dependent on the definition of privacy used
\citep{hotzpnas, kenny2023comment}. Further research and discussion
about the privacy-accuracy trade-off and the real-world privacy
protections offered by a formally private system such as TopDown are
warranted.

Finally, while our analysis shows how to rigorously evaluate the DAS
without access to the CEF, the Bureau has all of the necessary
information---including the CEF---to much more precisely assess the bias
and noise of the DAS than we are able to do here. Although the Bureau
has released some summary statistics about the error induced by TopDown
\citep{census2023may}, we believe that our independent assessment,
combined with a methodology for computing the bias and variance, is
useful for policy makers and researchers.

Ideally, the Bureau could produce margins of error for decennial census
statistics that hold on average across all geographies nationwide and/or
runs of the TopDown algorithm. At a minimum, the Bureau should release a
wide variety of error metrics including mean error and mean square error
of the TopDown counts for all on- and off-spine geographies in a
machine-readable format. These error metrics should also be broken out
by total population of each geography, since as Figure~\ref{fig-bias}
and Figure~\ref{fig-rmse} show, bias and variance can vary with total
population. Given our findings of error variability in swapped counts,
the Bureau should also release a similar set of metrics for runs of
their 2010 swapping algorithm under production settings. No error
metrics of any sort have been publicly reported for the 1990, 2000, and
2010 swapping algorithms, to our knowledge. For these metrics to be
useful to researchers, especially in areas with small populations or
small racial groups, they would need to be quite accurate. This means
that releasing such error metrics would require Census Bureau decision
makers to again weigh the benefits of such a release against the
potential privacy risks.

\nocite{nmfdocumentation}

\bibliography{references.bib}

\section*{Acknowledgments}
\thanks{We thank Ruobin Gong and an anonymous reviewer of the Alexander
and Diviya Magaro Peer Pre-Review Program for useful comments, and
Michael B. Hawes of the Census Bureau's Research and Methodology
Directorate for sharing information about various details of the Noisy
Measurement File.

\subsection*{Competing Interests}\label{competing-interests}
\addcontentsline{toc}{subsection}{Competing Interests}

All authors declare that they have no competing interests.

\subsection*{Data and materials
availability}\label{data-and-materials-availability}
\addcontentsline{toc}{subsection}{Data and materials availability}

All data, as well as code to reproduce the analysis, are are available
as part of the supplemental repository at
\url{https://doi.org/10.7910/DVN/TMIN3H}.}

\clearpage
\setcounter{section}{0}

\vspace*{1in}

{\LARGE\centering
\vspace*{-1.5em}Supplementary Materials for

\textbf{Evaluating Bias and Noise Induced by the U.S. Census Bureau's
Privacy Protection Methods}}

\vspace{1in}

\textbf{This PDF file includes:}

Supplementary Sections S1--S4

Figs. S1--S3

References (1--2)

\textbf{Other Supplementary Materials for this manuscript include the
following:}

Data and replication code, \url{https://doi.org/10.7910/DVN/TMIN3H}.

\renewcommand\thesection{S\arabic{section}}
\renewcommand\thefigure{S\arabic{figure}}
\renewcommand\thetable{S\arabic{table}}
\clearpage

\section{Mapping the NMF geographies to the standard Census
geographies}\label{sec-mapping}

To construct noisy measurements for standard census geographies, we
first map the geocode of the NMF to traditional GEOIDs. Queries for the
NMF are produced along a non-traditional spine that treat American
Indian / Alaska Native (\aian{}) and non-\aian{} areas as separate
components. This means that the NMF does not contain noisy measurements
for some of standard census ``spine'' geographies such as census block
groups and census tracts. Although this problem does not apply to a
minority of states that contain no \aian{} area (AR, DC, DE, IL, IN, KY,
MD, MO, NH, NJ, OH, PA, TN, VT, WV), most states and hence the census
geographies within them are split into \aian{} and non-\aian{} areas.
For the sake of completeness, this section describes in detail how we
mapped the NMF geographies to the standard Census geographies.

\subparagraph{Detailed Geocode--GEOID
correspondence.}\label{detailed-geocodegeoid-correspondence.}

An NMF geocode is a unique identifier for each geography. The NMF
contains basic documentation on the correspondence between geocodes and
GEOIDs \citep{nmfdocumentation}. Like in the usual decennial census
releases, the lowest level of geography is the census block, for which
the NMF geocodes are 31 digits long. All census geographies can be built
by aggregating census blocks. Most blocks are included in the TopDown
algorithm's process with the exception of the blocks that have zero
population. Every included block has a one-to-one mapping between
geocodes and traditional GEOIDs. Other geographies may or may not have a
unique mapping between them. This is in large part a function of if a
state or county has any \aian{} areas or not, where units are more
similar if there are no \aian{} areas.

An example of a valid geocode for a census block is
\texttt{0531000100011065300195010011010}. The first digit indicates
whether this block is on the \aian{} spine (\texttt{1}) or not
(\texttt{0}) \citep[item 14]{nmfdocumentation}. The final 16 digits is
related to the GEOID of the block \citep[item 15]{nmfdocumentation}.
Specifically, the GEOID for a block is the 16th to the 26th digit of the
geocode concatenated with the 28th to 31th digits. The 27th digit, which
refers to the traditional block group, is replicated in geocodes, though
it is (correctly) never replicated in traditional GEOIDs. As such, it
can be dropped to create traditional GEOIDs. Besides the first and final
16 digits, the other digits have substantive meanings. These are
explained in the table below.

\begin{longtable}[]{@{}
  >{\raggedright\arraybackslash}p{(\columnwidth - 6\tabcolsep) * \real{0.1000}}
  >{\raggedleft\arraybackslash}p{(\columnwidth - 6\tabcolsep) * \real{0.1000}}
  >{\raggedleft\arraybackslash}p{(\columnwidth - 6\tabcolsep) * \real{0.1000}}
  >{\raggedright\arraybackslash}p{(\columnwidth - 6\tabcolsep) * \real{0.5000}}@{}}
\caption{Explanation of digits of the NMF geocodes.}\tabularnewline
\toprule\noalign{}
\begin{minipage}[b]{\linewidth}\raggedright
Component
\end{minipage} & \begin{minipage}[b]{\linewidth}\raggedleft
Start
\end{minipage} & \begin{minipage}[b]{\linewidth}\raggedleft
End
\end{minipage} & \begin{minipage}[b]{\linewidth}\raggedright
Description
\end{minipage} \\
\midrule\noalign{}
\endfirsthead
\toprule\noalign{}
\begin{minipage}[b]{\linewidth}\raggedright
Component
\end{minipage} & \begin{minipage}[b]{\linewidth}\raggedleft
Start
\end{minipage} & \begin{minipage}[b]{\linewidth}\raggedleft
End
\end{minipage} & \begin{minipage}[b]{\linewidth}\raggedright
Description
\end{minipage} \\
\midrule\noalign{}
\endhead
\bottomrule\noalign{}
\endlastfoot
1 & 1 & 1 & \aian{} digit; 0 = non-\aian{} , 1 = \aian{} \\
2 & 2 & 3 & state FIPS code \\
3 & 4 & 5 & spine optimization code; 10 = on spine, county or below \\
4 & 6 & 8 & county FIPS \\
5 & 9 & 12 & tract FIPS-equivalent \\
6 & 13 & 15 & optimize block group FIPS-equivalent \\
7 & 16 & 17 & state FIPS (traditional GEOID component) \\
8 & 18 & 20 & county FIPS (traditional GEOID component) \\
9 & 21 & 26 & tract FIPS (traditional GEOID component) \\
10 & 27 & 27 & block group FIPS; repeat of first digit of component 10
(block FIPS) \\
11 & 28 & 31 & block FIPS (traditional GEOID component) \\
\end{longtable}

\subparagraph{Creating Crosswalks to Standard Census
Geographies.}\label{creating-crosswalks-to-standard-census-geographies.}

As discussed above, each standard census geography corresponds to a
collection of (non-overlapping) \aian{} and non-\aian{} pieces. A census
block is never split between \aian{} and non-\aian{} pieces. Instead, it
belongs to either the \aian{} spine or non-\aian{} spine. Spines ``do
not subdivide tabulation blocks'' \citep[item 12]{nmfdocumentation}.

A census block-group is a collection of census blocks and hence may be
split between the \aian{} and non-\aian{} spines. The same applies to a
census tract. Optimized block groups are special groups of blocks which
are only used in the TopDown algorithm. These block groups are designed
to better nest into off-spine geographies of interest, such as census
places. Traditional block groups do not necessarily optimize this, but
are used frequently in research, as the American Community Survey often
releases data at the block group level. These relationships are shown in
Figure 1 in the main text.

We use a simple algorithm to produce noisy measurements for standard
census geographies that are not on the NMF spines. Our algorithm is
designed to produce minimum-variance estimates for off-spine geographies
subject to the constraint that estimates be produced by summing
component geographies. This constraint is sufficient (though perhaps not
necessary) to ensure that resulting estimates from disjoint geographies
are statistically independent, a critical requirement for further
analysis.

There often exist more than one set of non-overlapping NMF geographies
that exactly correspond to any given standard census geography. To
minimize the variance of the resulting noisy measurements, we compose
the desired area from a small number of large non-overlapping NMF
geographies. This is because large NMF geographies have smaller
variances and adding more NMF geographies leads to a greater variance.

We start with the largest NMF geography that fits within a given target
geography, working through the NMF spine hierarchically from counties
down to census blocks. For example, for a given voting district (VTD),
we find all of the optimized block groups that are fully contained
within it (after checking both counties and tracts, which are generally
never fully contained or coterminous with VTDs). The noisy measurements
for these optimized block groups are used in the VTD estimate. We then
include the noisy measurements for the census blocks covering the
remaining areas of the VTD. The noisy measurements and their variances
for this VTD can be obtained by simply summing the corresponding noisy
measurements of the on-spine NMF geographies that are used to fill the
VTD.

Procedures may exist which produce smaller-variance estimates for a set
of geographies or a specific geography. One way to do so involves
subtracting small NMF geographies from larger ones to fill a standard
census geography. However, this approach (and other alternatives) will
tend to create dependencies between noisy measurements because the same
noisy measurement will be used to produce estimates for different
off-spine geographies. This significantly complicates subsequent
statistical analysis. As a result, we do not use this approach.

\section{Proofs of Propositions}\label{sec-proofs}

In this section, we provide proofs for all the propositions of this
section.

\subparagraph{Proposition 2.1.}\label{proposition-2.1.}

\begin{proof}
Since $\{\vb*\eta_\gamma\}_{\gamma\in\Gamma}$ are mutually independent across $\gamma\in\Gamma$, and $\vb*\eps^\nm_\gamma \coloneq B(\Sigma_\gamma)\vb*\eta_\gamma$ is a function of $\vb*\eta_\gamma$ alone ($B(\Sigma_\gamma)$ is fixed \textit{a priori}), $\{\vb*\eps^\nm_\gamma\}_{\gamma\in\Gamma}$ are therefore also mutually independent, proving (a).

Similarly, we must also have that $\{\vb*\eps^\nm_\gamma\}_{\gamma\in\Gamma}$ and $\{\vb*\eps^\sw_\gamma\}_{\gamma\in\Gamma}$, since each $\vb*\eta^\nm_\gamma$ is drawn independently, proving (b).
\end{proof}

\subparagraph{Proposition 2.2.}\label{proposition-2.2.}

\begin{proof}
The $\{\tilde{\vb*\eps}^\pp_\gamma\}_{\gamma\in\Gamma}$ are a function of $\{\tilde{\vb*\eta}_\gamma\}_{\gamma\in\Gamma}$ alone, and so are independent of $\{\vb*\eps^\nm_\gamma\}_{\gamma\in\Gamma}$, which are a function of $\{\vb*\eta_\gamma\}_{\gamma\in\Gamma}$ alone, proving (a).

Similarly, the $\{\tilde{\vb*\eps}^\pp_\gamma\}_{\gamma\in\Gamma}$ are independent of $\{\vb*\eps^\pp_\gamma\}_{\gamma\in\Gamma}$, which are a function of $\{\vb*\eta_\gamma\}_{\gamma\in\Gamma}$ alone, proving (b).
\end{proof}

\subparagraph{Proposition 2.3.}\label{proposition-2.3.}

\begin{proof}
The variance of $\hat{\tilde\mu}^\td_{\I s}$ is
\begin{align}
    \Var[\hat{\tilde\mu}^\td_{\I s}] 
    &= \frac{1}{|\I|^2}\qty(
    \sum_{\gamma\in\I} \Var[\widetilde Y^\pp_{\gamma s} - Y^\nm_{\gamma s}] + 
    \sum_{\mathclap{\gamma,\gamma^\prime\in\I,\ \gamma\neq\gamma^\prime}}
        \Cov(\widetilde Y^\pp_{\gamma s} - Y^\nm_{\gamma s}, \widetilde Y^\pp_{\gamma^\prime s} - Y^\nm_{\gamma^\prime s}) ) \\
    &= \frac{1}{|\I|^2}\qty(
        \sum_{\gamma\in\I} \qty(\Var[\tilde\eps^\pp_{\gamma s}] + \sigma^2_{\gamma s}) + 
        \sum_{\mathclap{\gamma,\gamma^\prime\in\I,\ \gamma\neq\gamma^\prime}}
        \Cov(\tilde\eps^\pp_{\gamma s}, \tilde\eps^\pp_{\gamma^\prime s}) ),
\end{align}
since by Proposition \ref{prop-indep-rerun}(a) $\{\vb*\eps^\nm_\gamma\}_{\gamma\in\Gamma}$ and $\{\vb*{\tilde\eps}^\pp_\gamma\}_{\gamma\in\Gamma}$ are independent, and by Proposition \ref{prop-indep-methods}(a) the $\{\vb*{\tilde\eps}^\nm_\gamma\}_{\gamma\in\Gamma}$ are mutually independent.

For any $\gamma,\gamma^\prime\in\I$, we have 
\begin{align*}
    \E[(Y^\pp_{\gamma s} - \widetilde Y^{\pp}_{\gamma s})(Y^\pp_{\gamma^\prime s} - 
    \widetilde Y^{\pp}_{\gamma^\prime s})]
    &= \E[(Y + \eps^\pp_{\gamma s} - Y - {\tilde\eps}^{\pp}_{\gamma s})
    (Y + \eps^\pp_{\gamma^\prime s} - Y - {\tilde\eps}^{\pp}_{\gamma^\prime s})] \\
    &= \E[(\eps^\pp_{\gamma s} - {\tilde\eps}^{\pp}_{\gamma s})
    (\eps^\pp_{\gamma^\prime s} - {\tilde\eps}^{\pp}_{\gamma^\prime s})] \\
    &= 2\Cov(\eps^\pp_{\gamma s}, \eps^\pp_{\gamma^\prime s}),
\end{align*}
where the final line follows because by Proposition \ref{prop-indep-rerun}(b), $\{\vb*\eps^\pp_\gamma\}_{\gamma\in\Gamma}$ and $\{\vb*{\tilde\eps}^{\pp}_\gamma\}_{\gamma\in\Gamma}$ are independent.

Then 
\begin{align*}
    \E\qty[\frac{1}{2}\qty(\sum_{\gamma\in\I} Y^\pp_{\gamma s} - \widetilde Y^\pp_{\gamma s})^2]
    &=  \frac{1}{2}\sum_{\gamma,\gamma^\prime\in\I}
        \E\qty[(Y^\pp_{\gamma s} - \widetilde Y^\pp_{\gamma^\prime s})
        (Y^\pp_{\gamma s} - \widetilde Y^\pp_{\gamma^\prime s})] \\
    &=  \frac{1}{2}\sum_{\gamma,\gamma^\prime\in\I} 2\Cov(\eps^\pp_{\gamma s}, \eps^\pp_{\gamma^\prime s}) \\
    &=  \sum_{\gamma\in\I} \Var[\eps^\pp_{\gamma s}]+
        \sum_{\mathclap{\gamma,\gamma^\prime\in\I,\ \gamma\neq\gamma^\prime}} 
        \Cov(\eps^\pp_{\gamma s}, \eps^\pp_{\gamma^\prime s}).
\end{align*}

Thus adding in $\sum_{\gamma\in\I}\sigma^2_{\gamma s}$ and dividing by $|\I|^2$, we find that \[
    \E[\widehat{V}(\hat{\tilde\mu}^\td_{\I s})] = \Var[\hat{\tilde\mu}^\td_{\I s}].
\]
\end{proof}

\subparagraph{Proposition 2.4.}\label{proposition-2.4.}

\begin{proof}
The variance of $\hat{\mu}^\sw_{\I s}$ is
\begin{align}
    \Var[\hat{\mu}^\sw_{\I s}] 
    &= \frac{1}{|\I|^2}\qty(
    \sum_{\gamma\in\I} \Var[Y^\sw_{\gamma s} - Y^\nm_{\gamma s}] + 
    \sum_{\mathclap{\gamma,\gamma^\prime\in\I,\ \gamma\neq\gamma^\prime}}
        \Cov(Y^\sw_{\gamma s} - Y^\nm_{\gamma s}, Y^\sw_{\gamma^\prime s} - Y^\nm_{\gamma^\prime s}) ) \\
    &= \frac{1}{|\I|^2}\qty(
        \sum_{\gamma\in\I} \qty(\Var[\eps^\sw_{\gamma s}] + \sigma^2_{\gamma s}) + 
        \sum_{\mathclap{\gamma,\gamma^\prime\in\I,\ \gamma\neq\gamma^\prime}}
        \Cov(\eps^\sw_{\gamma s}, \eps^\sw_{\gamma^\prime s}) ), \\
    &\le \frac{1}{|\I|^2}
        \sum_{\gamma\in\I} \qty(\Var[\eps^\sw_{\gamma s}] + \sigma^2_{\gamma s}),
\end{align}
with the final line following by assumption.

For any $\gamma\in\I$, 
\begin{align*}
    \E\qty[(Y^\sw_{\gamma s} - Y^\nm_{\gamma s})^2]
    &= \E\qty[(\eps^\sw_{\gamma s} - \eps^\nm_{\gamma s})^2] \\
    &= \E[{\eps^\sw_{\gamma s}}^2] + \sigma^2_{\gamma s} \\
    &\ge \Var[\eps^\sw_{\gamma s}] + \sigma^2_{\gamma s},
\end{align*}
where the third line follows because $\eps^\nm_{\gamma s}$ is zero-mean and independent of $\eps^\sw_{\gamma s}$ by Proposition \ref{prop-indep-methods}(b).

Combining these two results, by the linearity of expectation \[
    \E[\widehat{V}(\hat\mu^{\sw}_{\I s})] \ge \Var[\hat{\mu}^{\sw}_{\I s}].
\]
\end{proof}

\section{Average Bias and Root Mean Square Error for NH/PI, Other, and
Two or More
Races}\label{average-bias-and-root-mean-square-error-for-nhpi-other-and-two-or-more-races}

\FloatBarrier

We replicate Figures 4 and 5 in the main text for the remaining three
top-level race groups. These are Native Hawaiian and Pacific Islander
(NH/PI), some other race, and those who identify as two or more races.

\begin{figure}[t]

\centering{

\includegraphics[width=1\textwidth,height=\textheight]{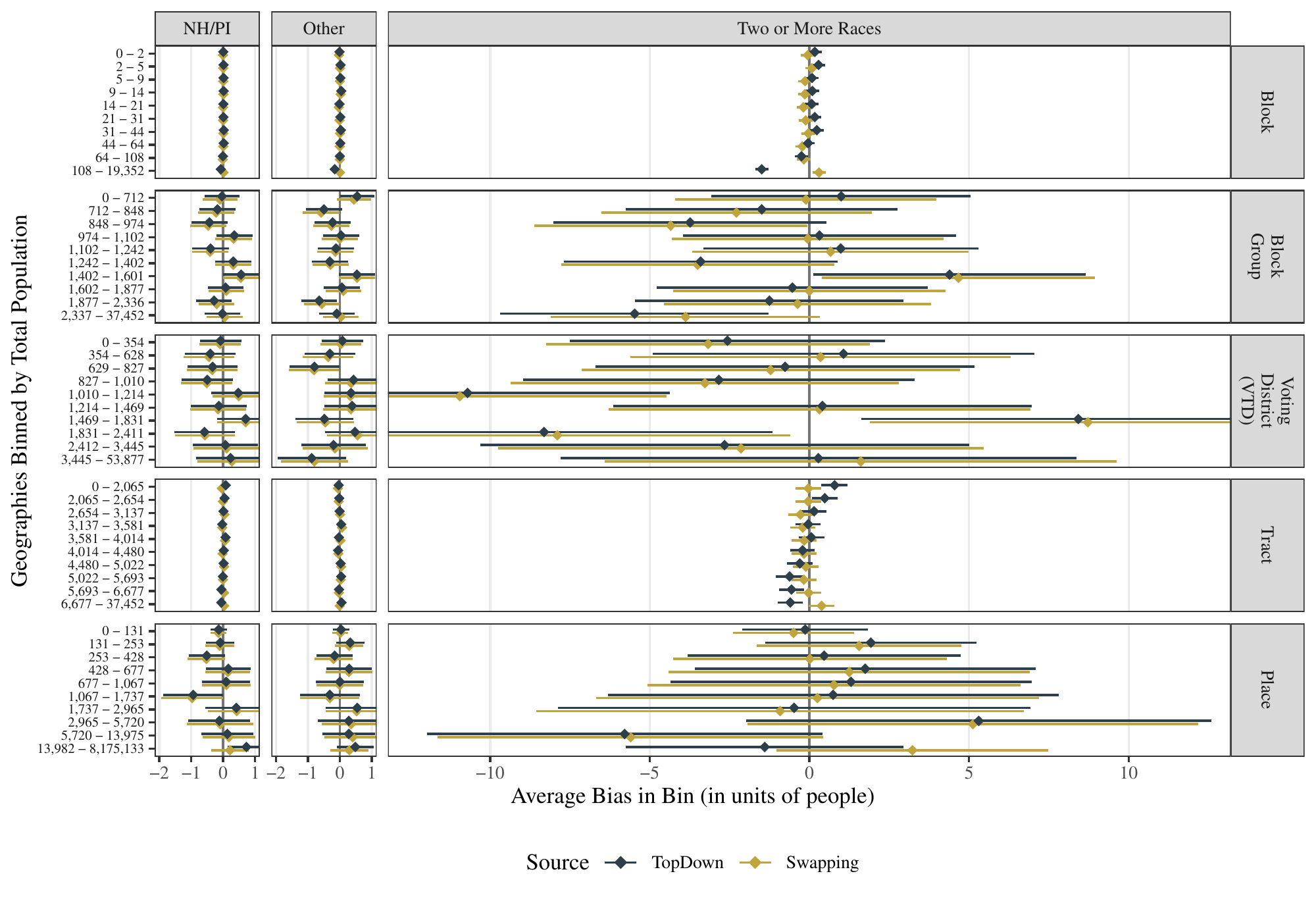}

}

\caption{\label{fig-app-bias}\textbf{Average bias for race/ethnicity
population counts at each geographic level by its total population.} The
figure estimates the average overcounting or undercounting in a group of
geographies, separately for five geographic levels and the remaining
three race/ethnicity groups. See ``Proposed Estimators'' in the main
text for estimators. Bins on the \emph{y}-axis are deciles of total
population of the geographic level Points show the estimated bias, and
lines show estimated 95\% confidence intervals.}

\end{figure}

As seen in Figure~\ref{fig-app-bias}, the average biases for NH/PI and
Other race groups are estimated to be small. This is similar to the
patterns for White, Black, Asian, and \aian{} racial groups. For members
of two or more race groups, the estimated average biases vary more. For
two or more race groups, higher population blocks and tracts tend to
display downward average bias, while the smallest blocks and tracts are,
on average, biased towards having more population. Estimates for
off-spine geographies are noisier and do not display a clear pattern of
average bias across populations.

\begin{figure}[t]

\centering{

\includegraphics[width=1\textwidth,height=\textheight]{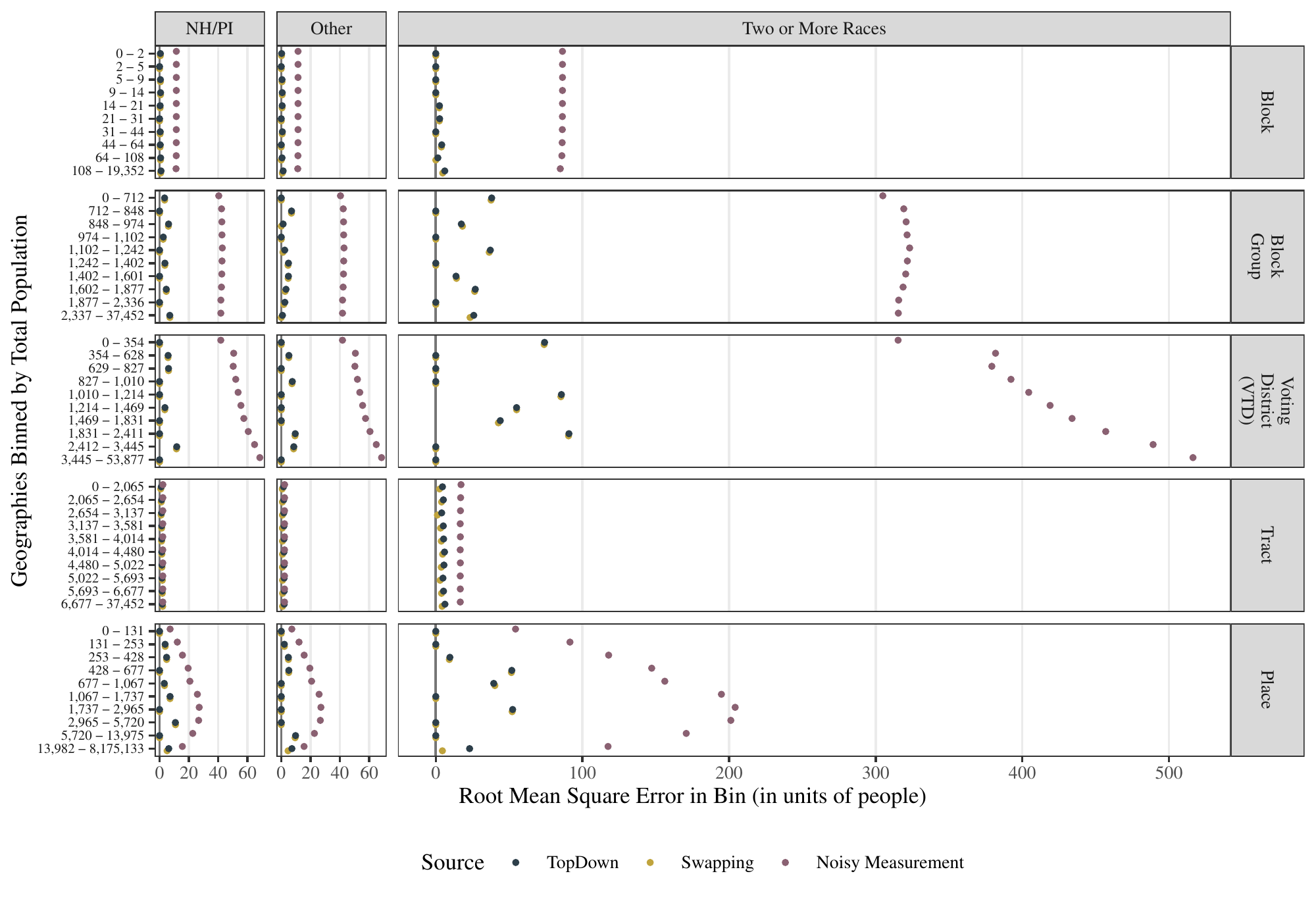}

}

\caption{\label{fig-app-rmse}\textbf{Estimated root mean square error of
race/ethnicity counts at each geographic level, by its total
population.} This estimates RMSE for a subset of geographies. See
``Proposed Estimators'' in the main text for estimators. The RMSE
estimates for Post-Processing and Swapping error are occasionally
negative, and these are imputed at 0. Bins on the \emph{y}-axis are
deciles by total population of the geographic level.}

\end{figure}

Figure~\ref{fig-app-rmse} similarly replicates the trends in Figure 5.
The RMSE for NH/PI and Other are smallest and relatively constant across
geography sizes for on-spine geographies. The RMSEs are largest for the
two or more race groups, likely following a similar logic to Hispanics,
where separate queries are used for this grouping. As in Figure 5, the
NMF is clearly the noisiest, with the post-processing step of the
TopDown algorithm reducing the variation to about the same scale as the
noise from swapping.

\section{Local Linear Regression with Measurement
Error}\label{sec-locallinear}

Figure 6 in the main text shows results of a local linear regression
where we regress the error in the counting of total population, which is
known exactly, on the percentage of people who are non-White. We use the
method proposed by \citet{fan2009JASA} to fit this regression based on
the implementation available in the R package \texttt{lpme}. We use 50
equally separated evaluation points \(\{0, 0.02, 0.04, \cdots, 1\}\)
along the values of percentage non-white, and a bandwidth of +/- 5
percentage points. The sequence of 50 fitted values from the models form
a local linear regression that approximates the true nonparametric
model. We fit this model for block groups, tracts, VTDs, and places. We
do not fit it to blocks because the values of the fraction non-White are
too concentrated (e.g., in values such as 1/3 or 1/2). This
concentration is due to the large left skew of the distribution of block
populations, where many blocks have near-zero populations.

This nonparametric regression approach is designed to account for
measurement error by incorporating the standard deviation of the
injected noise. As implemented in the \texttt{lpme} package, the method
only allows for the homoskedastic variance. Although not ideal, we use
the median variance of noisy measurements across geographies (rescaled
by the square of the total population for the fraction) as the input to
the model.

Additionally, the existing methodology assumes that each observation is
independent of the others. In fact, the TopDown errors are correlated
across geographies. Just as with OLS, we expect this correlation to
affect the variance of the nonparametric estimate, not the bias although
we have not rigorously proved this conjecture. In addition, statistical
results for the method we use are all asymptotic; here, we have a large
but fixed number of Census geographies and so the results should be
interpreted with some caution.

\begin{figure}[t]

\centering{

\includegraphics[width=1\textwidth,height=\textheight]{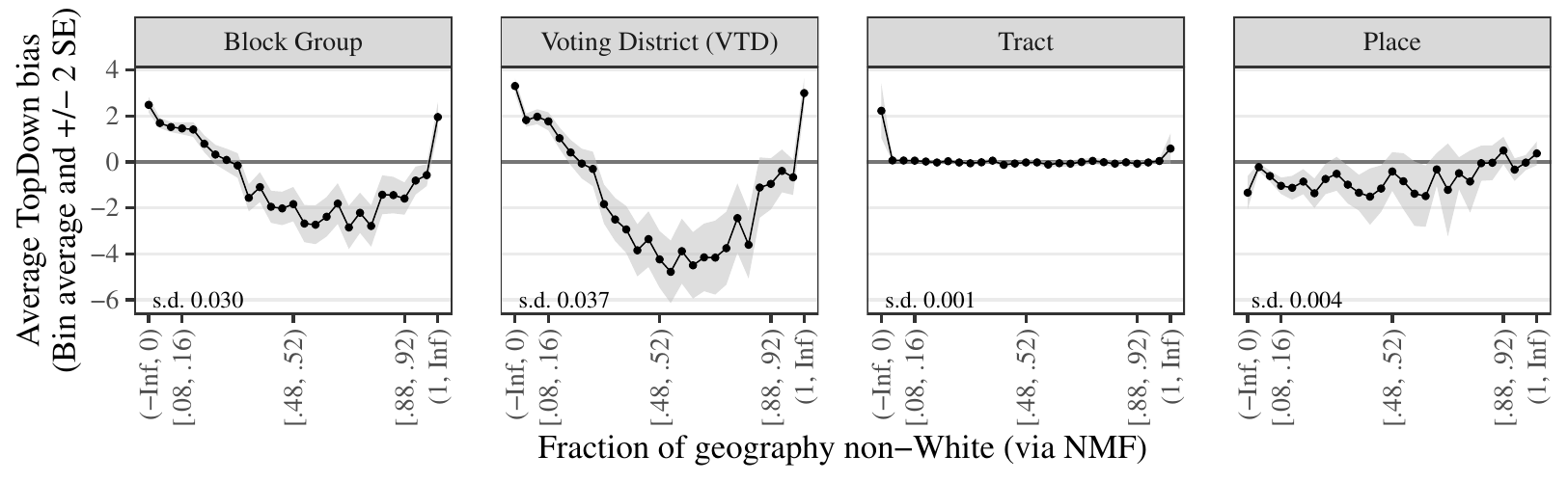}

}

\caption{\label{fig-lpme-alt}\textbf{Average bias in total population
count by percent non-white, with mutually exclusive bins.} Similar to
Figure 6, the figure estimates the degree of average TopDown
overcounting or undercounting by the racial demographics of the
geographic unit. We show the average error in each mutually exclusive
bin of the noisily measured percentage non-White. Bands show twice the
standard error around each average.}

\end{figure}

We also implemented an alternative approach to model this relationship
that enables some quantification of uncertainty.
Figure~\ref{fig-lpme-alt} uses mutually exclusive bins of percentage
non-White measure, instead of a moving average. We binned data between 0
and 100\% non-White by bins 4 percentage points wide. Negative values
and values above 100\% --- which arise due to the noisy measurement of
the non-White population --- are placed in their own bin. Together, for
each level of geography, we partition the units into 27 bins. We then
compute the simple average of the total population counting error of the
geographic units in each bin, with a standard error of the mean computed
by the standard deviation of the error divided by the square root of the
number of geographic units in the bin. This method cannot account for
the substantial measurement error in inherent in the percentage
non-White measure, but we can provide a rough measure of uncertainty.
The s.d. in Figure~\ref{fig-lpme-alt} represents the median of the
standard deviation inherent in the measure of percentage non-White for a
unit in each level. For example, the noisy measurement induces a noise
of about 3 percentage points for the median block group. The pattern of
local averages here mirrors the trend in Figure 6. Moreover, it shows
that the undercounting bias is in off-spine geographies is statistically
significant from zero.

\end{document}